\documentclass[12pt,preprint]{aastex}
\usepackage{graphicx}
\begin{document}

\title{A Search for Carbon-Chain-Rich Cores in Dark Clouds}
\author{Tomoya HIROTA, and Masatoshi OHISHI}
\affil {National Astronomical Observatory of Japan, Mitaka, Tokyo 181-8588, JAPAN }
\affil {tomoya.hirota@nao.ac.jp}
\author{Satoshi YAMAMOTO}
\affil {Department of Physics, The University of Tokyo, Bunkyo-ku, Tokyo 113-0033, JAPAN }

\begin{abstract}

We present results of a survey of CCS, HC$_{3}$N, and HC$_{5}$N 
toward 40 dark cloud cores to search for 
"Carbon-Chain--Producing Regions(CCPRs)", where carbon-chain molecules 
are extremely abundant relative to NH$_{3}$, 
as in L1495B, L1521B, L1521E, and the cyanopolyyne peak of TMC-1. 
We have mainly observed toward cores where the 
NH$_{3}$ lines are weak, not detected, or not observed in 
previous surveys, and the CCS, HC$_{3}$N, and HC$_{5}$N lines 
have been detected toward 17, 17, and 5 sources, respectively. 
Among them, we have found a CCPR, L492, 
and its possible candidates, L1517D, L530D, L1147, and L1172B. 
They all show low abundance ratios of [NH$_{3}$]/[CCS] 
(hereafter called the NH$_{3}$/CCS ratio) 
indicating the chemical youth. 
Combining our results with those of previous surveys, 
we have found a significant variation of the NH$_{3}$/CCS ratio 
among dark cloud cores and among molecular cloud complexes. 
Such a variation is also suggested 
by the detection rates of carbon-chain molecules. 
For instance, the NH$_{3}$/CCS ratios are higher and the detection rates 
of carbon-chain molecules are lower in the Ophiuchus cores than in the Taurus cores. 
An origin of these systematic abundance variation is discussed in terms of the 
difference in the evolutionary stage or the contraction timescale. 
We have also identified a carbon-chain-rich star-forming core, L483, 
where intense HC$_{3}$N and HC$_{5}$N lines are detected. 
This is a possible 
candidate for a core with "Warm Carbon-Chain Chemistry". 
\end{abstract}


\section{Introduction}
  
One of the outstanding characteristics of dark cloud cores, which are 
formation sites of low-mass stars, is high abundances of carbon-chain 
molecules. \citet{suzuki1992} carried out survey 
observations of CCS, HC$_{3}$N, HC$_{5}$N, and NH$_{3}$ 
toward 49 dense cores, and found that the abundance of CCS is anticorrelated 
with that of NH$_{3}$, which is interpreted in terms of chemical evolution 
of dark cloud cores; CCS and NH$_{3}$ are abundant in early and late stages of 
chemical evolution, respectively \citep{suzuki1992}. 
Similar survey observations were carried out for 
dense molecular cloud cores \citep{fuente1990, benson1998, lai2000}, 
Bok globules and Herbig Ae/Be stars \citep{scappini1996}, 
translucent clouds \citep{turner1998}, 
low-mass young stellar objects associated with H$_{2}$O masers 
\citep{degregorio-monsalvo2006}, and infrared dark clouds \citep{sakai2008}, 
as well as individual clouds such as the Pipe Nebula \citep{rathborne2008}, 
the Perseus molecular cloud \citep{rosolowsky2008}, and the Orion 
molecular cloud \citep{tatematsu2008}, 
revealing a relationship between physical and chemical evolutionary stages 
of molecular cloud cores. 
A number of detailed studies on the dark cloud cores using the CCS 
lines were also conducted \citep[e.g.,][]{hirahara1992, 
velusamy1995, kuiper1996, ohashi1999, lai2003}. 
These results clearly demonstrated that the distribution of CCS is 
anticorrelated with those of NH$_{3}$ and N$_{2}$H$^{+}$. 
In particular, dynamically evolved cores such as L1498 and L1544, 
which are gravitationally infalling in the central part of the cores, show 
a donut-like distribution of CCS due to its depletion in gas phase, 
while the NH$_{3}$ and N$_{2}$H$^{+}$ distributions are centrally condensed 
\citep{kuiper1996, ohashi1999}. 
Such a chemically differentiated feature is successfully interpreted in 
terms of chemical and dynamical evolution of dark cloud cores 
through recent detailed theoretical studies 
\citep[e.g.,][]{bergin1997, aikawa2001, aikawa2003, aikawa2005}. 

In the CCS survey by \citet{suzuki1992}, 
four "Carbon-Chain--Producing Regions" (CCPRs) were identified: 
L1495B, L1521B, L1521E, and the cyanopolyyne peak of TMC-1. 
In CCPRs, the CCS and other carbon-chain molecules are extremely abundant while 
NH$_{3}$ is deficient, suggesting  
that they are chemically less evolved \citep{suzuki1992}. 
Recent comprehensive observational studies on CCPRs 
revealed that they are dynamically younger than other starless cores 
\citep{hirota2002, hirota2004, tafalla2004}. 
Thus, detailed studies on such sources are of great importance to 
understand initial conditions of low-mass star formation processes. 

However, only four CCPRs have been identified 
to date, all of which are in the Taurus molecular cloud. 
This is partly because most of the previous studies on dark cloud cores 
were based on the survey of the NH$_{3}$ lines 
\citep{benson1989} and the samples were primarily chosen from 
the NH$_{3}$ cores \citep[e.g.,][]{benson1998}. 
In order to establish the chemical evolutionary scheme of dark cloud cores 
on a statistical basis, we need to know how frequently CCPRs show up.  

With this in mind, we carried out a survey of the CCS, HC$_{3}$N, and 
HC$_{5}$N lines toward dark cloud cores where the 
NH$_{3}$ lines are weak, not detected, or not observed in the 
previous surveys, considering 
that the abundances of carbon-chain molecules are 
anticorrelated with that of NH$_{3}$ \citep{suzuki1992}. 
As a result, we found a new candidate for a CCPR, L492 
in the Aquila rift, whose details were already published \citep{hirota2006}. 
In addition, we have reanalyzed previous observations of 
the CCS and NH$_{3}$ lines, part of which were published \citep[e.g.,][]{hirota2001}. 
All the data are used for the statistical analysis of the chemical 
evolution of dark cloud cores. 
Although our study is not based on unbiased survey observations, 
it would be useful to establish the chemical evolutionary scenario. 

\section{Observations}

Observations were conducted with the 45 m radio telescope at Nobeyama Radio 
Observatory (NRO) in 2002 March and 2004 April. 
The observed lines are summarized in Table \ref{tab-obsedf}. 
The CCS ($J_{N}$=4$_{3}$-3$_{2}$), 
HC$_{3}$N ($J$=5-4), and HC$_{5}$N ($J$=17-16) lines in the 45 GHz region 
were simultaneously observed with an SIS mixer receiver whose 
system temperature was 200-300 K. 
The main-beam efficiency ($\eta_{mb}$) and the half-power beam width 
were 0.7 and 37\arcsec, respectively. 
In addition, we observed the CCS and NH$_{3}$ lines in the 22-23 GHz region 
toward selected sources, L1400B, L1517D, L492, L429-2, L483, L530D, L1147, and L1172B, 
in order to search for NH$_{3}$ toward the CCS peak. 
These lines were observed simultaneously with a HEMT receiver whose system 
temperature was about 200 K. 
The main-beam efficiency and the beam size were 0.8 and 74\arcsec, respectively. 
Acousto-optical radio spectrometers with the frequency resolution of 37 kHz were 
used as the backend. Pointing was checked by observing nearby SiO maser sources 
every two hours, and the pointing accuracy was estimated to be better than 5\arcsec rms. 
Observations in 2002 March were performed with the position-switching mode 
with the off position of 10\arcmin \ in the azimuth direction. 
On the other hand, the frequency-switching mode was 
employed for the observations in 2004 April, with the frequency throw of 0.4~MHz 
for the off-signal spectrum. 

Observed sources are summarized in Table \ref{tab-source}. 
We selected these sources mainly from the list of the NH$_{3}$ survey by \citet{benson1989}, 
where the NH$_{3}$ lines are not detected or not intense 
in spite of the existence of a dense core traced by the C$^{18}$O line \citep{myerslinkebenson1983}. 
Furthermore, we selected the sources where the NH$_{3}$ lines were not 
observed by \citet{benson1989}, whereas the CS lines were detected \citep{lee2001}. 
In total, we observed 32 dark cloud cores, 
among which five sources contain {\it{IRAS}} point sources and the others are starless. 
They include five sources, L1523, L183(N), L1719B, B68, and L483 
where the CCS lines were previously 
observed \citep{suzuki1992, lai2003, benson1998}. 
Two sources, L1495C and L1172D, were later identified 
to be previously known sources, L1495N and L1172A, respectively. 

\begin{figure*}[htb]
\includegraphics[width=8cm]{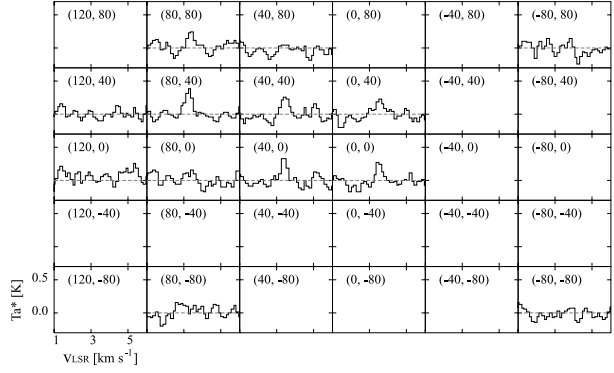}
\caption{Example of the profile map of the CCS lines in L530D. 
\label{fig-profile}  }
\end{figure*}

For most of the sources, we made five-point mapping observations 
around the reference positions, 
(0\arcsec, 0\arcsec), (+80\arcsec, +80\arcsec), (+80\arcsec, -80\arcsec), 
(-80\arcsec, +80\arcsec), and (-80\arcsec, -80\arcsec), in order not to 
miss the spectral lines, because the peak position of the core might often 
be shifted from the reference position by more than the beam size of 37\arcsec. 
For cores where at least one of the lines was detected at any position, 
we further searched for the positions with stronger intensity with 
a grid spacing of 40\arcsec. 
An example of the profile map obtained is shown in Figure \ref{fig-profile}. 
It should be noted that cores smaller than about 80\arcsec \ in diameter 
might be missed from detection in our observations due to the coarse-sampling observations. 

In addition, we analyzed results of our previous survey of 
CCS, HC$_{3}$N, and HC$_{5}$N toward eight sources, 
L1527, L1512, L183(S), L1696A, L1689N, L1157, L1155C, and L1228, as listed in 
Table \ref{tab-source-app}. 
For these sources, the results for CCS were reported by \citet{hirota2001}. 
Observations were made in 1998 and 1999 with the 45 m radio telescope at NRO. 
Details of the observations were similar to those of the sessions in 2002 and 2004. 
However, we observed these sources only toward the reference positions, except for 
L1512 where we carried out mapping observations of CCS and HC$_{3}$N. 
We observed the CCS and HC$_{3}$N lines toward all of the eight sources 
while HC$_{5}$N was observed only toward L1527. 

In total, we finally obtained 40 samples of dark cloud cores both 
with and without the {\it{IRAS}} point sources. 

\section{Results}

Examples of the observed spectra are shown in Figures \ref{fig-l483} and \ref{fig-sp}. 
We detected both the CCS and HC$_{3}$N lines toward 17 out of 40 samples, 
and detected the HC$_{5}$N lines toward five sources out of 
35 samples. Among our sources, detection of the NH$_{3}$ lines 
was reported for 24 sources out of 39 sources, mainly by \citet{benson1989} and 
the present study. All of these detections indicate that 
there exist dense cores with the H$_{2}$ densities of the order of 
10$^{4}$ cm$^{-3}$ \citep{benson1989, suzuki1992}. 
On the other hand, none of the CCS, HC$_{3}$N, HC$_{5}$N, and NH$_{3}$ lines 
were detected toward 14 out of our 40 samples 
probably due to insufficient sensitivity of our observations, 
and/or positional offset from the intensity peak. 
Some of these cores might have densities less than the critical 
densities of the observed lines, $\sim10^{4}$~cm$^{-3}$. 

A few of them show intense spectra of carbon-chain molecules. 
In particular, we found remarkably intense spectra of CCS, HC$_{3}$N, and HC$_{5}$N 
lines toward L492. Brightness temperatures of these lines toward L492 were 
comparable to those toward the known CCPRs such as L1495B, L1521B, 
L1521E, and the cyanopolyyne peak of TMC-1 
\citep[e.g.,][]{suzuki1992}. This result strongly suggests that L492 
would be extremely rich in carbon-chain molecules, which was confirmed by follow-up 
observations \citep{hirota2006}. 

\subsection{Abundances of the Observed Molecules}

For all of the detected lines, peak antenna temperature, line width, 
and LSR velocity are derived by fitting the Gaussian profile to each 
spectrum, which are summarized in Table \ref{tab-obsccs}. 
For the NH$_{3}$ lines toward L429-1 and L483, we derived the LSR velocities 
and the line widths assuming that they are common for all the hyperfine components. 
Because of the insufficient spectral resolution of 37 kHz, 
which is larger than most of the hyperfine splittings due to the H nuclei 
in the NH$_{3}$ (1,1) line, 
we fitted only the five major spectral components as labeled in Table \ref{tab-obsedf} 
\citep{ungerechts1980}. 
The line widths and LSR velocities are slightly 
different from molecule to molecule for some sources. This is 
partly due to the hyperfine structure of the HC$_{3}$N and NH$_{3}$ lines, 
whose line widths tend to be broader than those of the other molecules. 

Using these line parameters, column densities of CCS, HC$_{3}$N, HC$_{5}$N, and 
NH$_{3}$ were calculated by the similar method described by \citet{suzuki1992}. 
In addition, we calculated column densities of NH$_{3}$ for cores 
that we did not observe, using the results of \citet{benson1989} if they 
are available. 
The dipole moments of CCS, HC$_{3}$N, HC$_{5}$N, and NH$_{3}$ were assumed 
to be 2.81, 3.72, 4.33, and 1.46 Debye, respectively 
\citep{murakami1990, lafferty1978, alexander1976, cohen1974}. 
The LTE condition was assumed, and the excitation temperatures 
were fixed to 5~K for CCS and 6.5~K for HC$_{3}$N, HC$_{5}$N, and NH$_{3}$. 
In this paper, we do not use the results of $J_{N}$=$2_{1}$-$1_{0}$ lines of CCS, 
because the beam size at the 22 GHz band, 74\arcsec, is almost comparable to 
the typical size of the observed dark cloud cores as shown in Figure \ref{fig-maps}, 
and the derived column densities may be affected 
due to the unknown source coupling efficiency. 

The derived column densities are summarized in Table \ref{tab-collte}. 
For L492, line parameters and column densities are taken from \citet{hirota2006}. 
Calculated optical depths,$\tau$, of the observed lines were 
0.17-3.4, 0.07-8.8, and 0.15-0.65 for the CCS, HC$_{3}$N, and HC$_{5}$N lines, 
respectively. The CCS and HC$_{5}$N lines were found to be optically thin ($\tau<1$), 
except for the CCS line toward L492 \citep[$\tau=3.4$; ][]{hirota2006}. 
On the other hand, the HC$_{3}$N lines are optically thick toward 
L1527 ($\tau=2.4$), L1512 ($\tau=1.5$), B68 ($\tau=1.0$), 
L492 \citep[$\tau=8.8$; ][]{hirota2006}, and L483 ($\tau=1.5$). 
For L492, the optical depth and excitation temperatures were derived 
from the intensity ratio of the hyperfine components 
of the HC$_{3}$N line \citep{hirota2006}. 

In order to estimate uncertainties in the derived column densities, 
we calculated the column densities of CCS, HC$_{3}$N, and HC$_{5}$N 
by varying the assumed excitation temperature of $\pm$1~K. 
We found that the uncertainties in the derived column densities were 
typically 10\%-30\%. 
In addition, we compared column densities of some of the sources 
where the CCS lines were observed previously. 
We found that the column densities of CCS toward these sources differ 
by a factor of 0.4-3 mostly due to the difference in the observed positions, 
as in the case of L1512 and B68, for instance. 
Considering these results, we estimate that the uncertainties 
in the derived molecular column densities are typically about 30\%-40\%, 
while it could be as high as a factor of 3 in the worst case. 

\subsection{Distributions of the Observed Molecules}

For several sources where the CCS and HC$_{3}$N lines are intense, we obtained 
integrated intensity maps as shown in Figure \ref{fig-maps}. 
Detailed descriptions for individual sources are summarized in Appendix \ref{sec-individual}. 
The CCS and HC$_{3}$N maps show similar morphology \citep{hirahara1992, tafalla2006}, 
although some of them show slight differences in their sizes and peak positions. 
It is well known that CCS and some carbon-chain molecules are depleted at the 
central part of a starless dense core, and their distributions show 
a central hole \citep{velusamy1995, kuiper1996, ohashi1999, lai2000, lai2003}. 
We can see such an apparent hole only in the CCS map of L1172D, possibly due 
to the lack of significant depletion of CCS and HC$_{3}$N in other mapped sources. 
However, it is also likely that we could not trace such a small structure in the cores 
because of the large beam size and the coarse grid spacing of our mapping observations, 40\arcsec, 
which are comparable to or slightly larger than the expected size of the hole. 
In fact, the central hole of CCS in B335 was not detected by the single dish 
observation, but was first imaged by the interferometric observation using 
the VLA \citep{velusamy1995} with the spatial resolution of 12\arcsec. 
Further high-resolution observations would settle this issue. 

\section{Discussions}

\subsection{Possible Candidates for CCPRs}

In the present survey, we observed the CCS, HC$_{3}$N, and HC$_{5}$N lines 
toward 40 dark cloud cores, and found several dense cores with high column 
densities of carbon-chain molecules and/or extremely low NH$_{3}$/CCS ratios. 
Here we tentatively define the CCPR as the core with 
the NH$_{3}$/CCS ratio of less than 10, which can easily be 
distinguished from other cores in Table \ref{tab-collte} and 
Figure \ref{fig-nh3ccs} as discussed later. 
Based on this criterion, we can identify five new candidates for 
CCPRs, L492, L1517D, L530D, L1147, and the northeast clump of L1172B, as well as 
four known CCPRs, L1495B, L1521B, L1521E, and the cyanopolyyne 
peak of TMC-1 \citep{suzuki1992}. 
As briefly described in the Appendix \ref{sec-individual}, 
the chemical compositions of the newly found sources show 
a mutual variation, and are different from originally identified CCPRs. 
Here we classify the newly found candidates 
into two groups and summarize their properties below. 

One of the most outstanding source is L492, where 
the carbon-chain molecules have high column densities comparable 
to the known CCPRs. 
The high abundances of carbon-chain molecules, 
the lack of significant depletion, and low deuterium fractionation 
all suggest chemically less evolved nature of L492 
\citep[][and Appendix \ref{sec-l492}]{hirota2006}. 
Contrary to the other CCPRs, 
the NH$_{3}$ lines are also detected in L492 with 
moderate intensities. Furthermore, L492 is thought to be dynamically 
evolved according to a strong infalling signature found 
in the spectral lines \citep{lee2001}. 
Detailed discussions in \citet{hirota2006} concluded that L492 
is chemically and dynamically evolved than other CCPRs, 
while younger than prestellar cores such as 
L1498 and L1544 \citep{kuiper1996, ohashi1999}. 
This order does not always mean the absolute age, 
because timescales of contraction, gas phase chemistry, and depletion would be 
different from source to source. 

The other group of the sources with remarkably low NH$_{3}$/CCS ratios 
are the starless cores L1517D, L530D, and L1147, 
as well as the northeast clump of L1172B. 
In these sources, the NH$_{3}$/CCS ratios are comparable to those 
of other known CCPRs, although column densities of both of 
CCS and NH$_{3}$ are generally lower 
than the known CCPRs as can be seen in Table \ref{tab-collte}. 
Therefore, they are possible candidates for CCPRs. 
In the case of L1147, the weak NH$_{3}$ line is detected toward the dust continuum 
peak, and the CCS peak coincides with the same position 
(see Appendix \ref{sec-l1147}), 
which results in the low NH$_{3}$/CCS ratio toward the central part of the 
dense core as in the case of L492. However, the column densities 
of both CCS and NH$_{3}$ in L1147 are one order of magnitude lower than those of L492. 
Because the dust continuum emission of L1147 is weaker than 
the other cores \citep{kirk2005}, it is also likely that the column density of 
H$_{2}$ is also less than L492. 
On the other hand, in L1517D and L530D, the peak positions of CCS are located 
at the edge of dense cores, which would be a reason for the lower column densities 
of the observed molecules 
(see Appendices \ref{sec-l1517d} and \ref{sec-l530d}). 
Therefore, it cannot be ruled out the possibility 
that the CCS line in L1517D and L530D would trace the outer part of 
the dense core, as can be seen in L1498 and L1544 \citep{kuiper1996, ohashi1999}. 
For L1172B, dust continuum maps are not 
available in our knowledge, and hence, we will not discuss further the relationships 
between the CCS, HC$_{3}$N, NH$_{3}$, and dust continuum emission 
(see Appendix \ref{sec-l1172b}). 

In summary, L492 has been confirmed as 
a definitive CCPR that is found for the first time 
outside the Taurus molecular cloud, and the others are 
possible candidates for CCPRs, for which 
further high resolution mapping observations of 
carbon-chain molecules and NH$_{3}$, together with the dust continuum emission, 
are necessary for confirmation. 

\subsection{Detection Rate of the Carbon-Chain Molecules and NH$_{3}$}

According to our results, overall detection rates of the observed molecular lines 
are 43\% (17/40), 43\% (17/40), 14\%(5/35), and 62\%(24/39) for 
CCS, HC$_{3}$N, HC$_{5}$N, and NH$_{3}$, respectively, which are slightly 
lower than those of \citet{suzuki1992}, 55\%, 67\%, 31\%, and 80\%, respectively, 
as listed in Table \ref{tab-detection}. 
Although we selectively observed the cores with low abundance of NH$_{3}$, 
based on the anticorrelation between carbon-chain molecules and NH$_{3}$ 
\citep{suzuki1992}, 
the detection rates of carbon-chain molecules are not so high as expected. 
Most of the observed cores with weak NH$_{3}$ emission, 
except for five newly found candidates as mentioned in the previous section, 
are turned out to have "normal" NH$_{3}$/CCS ratios and hence, 
would be just low column density objects. 
As mentioned previously, none of the observed lines were detected toward 
some of our samples probably due to offsets from the real emission peak. 
Otherwise, they would have less densities so that the NH$_{3}$ lines along 
with those of carbon-chain molecules are hardly excited. 
These results mean that the CCPRs are intrinsically rare even in the Taurus region 
where four CCPRs have already been identified . 

As discussed in the previous section, 
we calculated the column densities of CCS, HC$_{3}$N, HC$_{5}$N, 
and NH$_{3}$ by the similar method described by \citet{suzuki1992}. 
This enables us to discuss statistically about the carbon-chain chemistry 
in dark cloud cores by combining our results with those of 
previous surveys \citep{suzuki1992, benson1989, benson1998, hirota2001} 
supplemented with several individual observations 
\citep[e.g.,][]{hirota2002, hirota2004}, which were made by almost 
the same method of observations and data analysis. 
Details of our samples are discussed in the Appendix \ref{sec-appendall}. 
We compiled a sample of nearby dark cloud cores consisting of 90 sources in total 
for the further statistical studies. The total detection rates are 
53\%, 55\%, 23\%, and 74\% for CCS, HC$_{3}$N, HC$_{5}$N, and NH$_{3}$, respectively. 

As noted in the Introduction, the CCPRs have so far been detected 
only in the Taurus molecular cloud \citep{suzuki1992}. 
This result would suggest a difference in the overall chemical compositions 
from cloud to cloud. 
Hence, it is worthy to compare the detection rates of carbon-chain molecules 
between the Taurus and other regions in order to investigate whether 
such a chemical abundance variation occurs. 
Because the numbers of available samples are sufficient only for the Taurus and 
Ophiuchus cores ($\sim$30 each), we here summarize the results for these two regions 
in Table \ref{tab-detection}. 
It is easily found that the detection rates of carbon-chain 
molecules are significantly lower in the Ophiuchus cores 
than in the Taurus cores. 
A ratio of the number of the CCS cores relative to that of the NH$_{3}$ cores 
in the Taurus region is three times higher than that of the Ophiuchus region, 
as listed in Table \ref{tab-detection}. 
Since our sample was not prepared in an unbiased way, 
these results are not statistical in a strict sense. Nevertheless, we can 
extract `statistical' trends from the results because the sample involves 
a number of sources with and without star-forming activities. 
Therefore, it is very likely that carbon-chain molecules are generally 
less abundant in the Ophiuchus cores than in the Taurus cores. 

We also compared our results with those of recent complete survey of 
CCS and NH$_{3}$ in the Pipe Nebula \citep{rathborne2008} and the Perseus molecular 
cloud \citep{rosolowsky2008}, as listed in Table \ref{tab-detection}. 
Detection rates of the observed molecules are different from region to region. 
In the Pipe Nebula, the detection rate of CCS is lower than that 
in the Taurus region and is 
comparable to that in the Ophiuchus region, while the detection rate of NH$_{3}$ is 
almost the same as those in the other regions. 
On the other hand, the detection rate of CCS in the Perseus region 
is rather close to that in the Taurus region, 
and the detection rate of NH$_{3}$ in the Perseus 
region is the highest among our four sampled regions. 
Although the methods of the surveys are different from ours in various ways, 
such as observed transitions, sensitivities, spectral and spatial resolutions, 
methods of data analysis, and definition of star formation activities 
(association of {\it{IRAS}} and/or {\it{Spitzer}} sources), 
the possible difference in the ratio of the number of the CCS cores 
relative to that of the NH$_{3}$ cores can be seen in Table \ref{tab-detection}; 
the ratio in the Taurus region (90\%) is nearly 3, 2, and 1.5 times 
higher than that in the Ophiuchus region, the Pipe Nebula, and the Perseus 
region, respectively, which would be indicative of difference in 
an average chemical age of cores in each region, 
as discussed later \citep{suzuki1992}. 

\subsection{Variation in the NH$_{3}$/CCS Ratio}

In order to investigate a source-to-source variation in molecular abundances, 
we focus on the abundance ratio of NH$_{3}$/CCS, which is recognized as a 
useful probe of the chemical evolutionary stage of dark cloud cores 
\citep{suzuki1992}. Following the discussion by \citet{ohishi1998}, 
we investigate the relationship between the CCS abundance and the NH$_{3}$/CCS 
ratio in Figure \ref{fig-nh3ccs}. 
We plotted the sources where either the CCS or NH$_{3}$ line is detected 
(Table \ref{tab-summary}). 
Because of the lack of a complete data set for the column densities of H$_{2}$ 
for all the cores, we plot the column densities of CCS instead of 
the fractional abundance of CCS relative to H$_{2}$. Therefore, a scatter in 
the column density of CCS is not only due to the variation in 
the fractional abundance of CCS relative to H$_{2}$, but is also attributed to 
the difference in the column density of H$_{2}$. 

In Figure \ref{fig-nh3ccs}, 
we can find CCPRs (L1495B, L1521B, L1521E, the cyanopolyyne peak of TMC-1, and L492) 
at the upper-left part, and the CCPR candidates 
(L1517D, L530D, L1147, and the (-40\arcsec, -40\arcsec) and 
(+40\arcsec, +40\arcsec) positions in L1172B) at the middle-left part. 
We stress that there is no carbon-chain-rich cores in the Ophiuchus region 
as depicted in Figure \ref{fig-nh3ccs}(d), 
even though we have almost the same number of samples as in the Taurus region. 
Such a systematic abundance variation from cloud to cloud has been 
found in the deuterium fractionation ratios of DNC/HN$^{13}$C \citep{hirota2001} 
and N$_{2}$D$^{+}$/N$_{2}$H$^{+}$ \citep{crapsi2005}. 

We again refer to the results of the Pipe Nebula \citep{rathborne2008} 
and the Perseus molecular cloud \citep{rosolowsky2008}. 
It can be found that there is no carbon-chain-rich core in the Perseus 
region in spite of the completeness of their survey, as shown in 
Figure \ref{fig-nh3ccs-p}. 
On the other hand, we find several candidates for CCPRs 
in the Pipe nebula \citep{rathborne2008}, where the NH$_{3}$/CCS ratios are less than 10. 
In the Pipe nebula, there exists a dense core \citep[core 37 of][]{rathborne2008} 
where the CCS line is detected while NH$_{3}$ is not. 
Because of the lack of the NH$_{3}$ data, it is not plotted in Figure \ref{fig-nh3ccs-p}(b). 
The column density of CCS toward this core is derived to be 
2.0$\times$10$^{12}$~cm$^{-2}$ \citep{rathborne2008}. 
Although this value is much lower than those of the known CCPRs, 
it would be a possible candidate of a CCPR. 
This result would suggest that the cores in the Pipe nebula are generally 
in the early stages of both chemical and dynamical evolution \citep{rathborne2008}. 

Although the above discussions are not based on unbiased, complete, and uniform surveys, 
the apparent region-to-region variation of the chemical composition would be 
indicative of difference in chemical and/or physical properties of each region. 
In fact, chemical and dynamical model calculations demonstrate that 
molecular abundances and distributions could reflect the physical properties 
and their initial conditions of the cores such as the density, degree of depletion, 
and velocity structure of the cores \citep{aikawa2001, aikawa2003, aikawa2005}. 
Observational evidences for the variations of physical properties are 
reported by \citet{jijina1999} based on the database of a survey of the NH$_{3}$ 
lines. Such a variation of the core properties would affect not only chemistry but 
also star-formation processes, which results in either isolated or clustered 
mode, and low-mass or high-mass star-formation. 

\subsection{Possible Origin of Chemical Abundance Variation}

In the above sections, we recognize significant source-to-source chemical abundance variation, 
in particular between the Taurus and Ophiuchus regions. 
Here we discuss two possible scenarios which can explain qualitatively the systematic 
difference in the molecular abundances between the Taurus cores and Ophiuchus ones. 

\subsubsection{Difference in Ages of the Cores}

Most simply, the difference in the chemical abundances between the Taurus 
and Ophiuchus regions can be interpreted 
in terms of the different age of the Taurus molecular cloud 
complex itself from that of Ophiuchus. 
Due to its younger chemical evolutionary stage of the Taurus region 
than in Ophiuchus, more young cores which have been just formed in the cloud 
complex could still remain in the Taurus region than in Ophiuchus. 

One would think that if CCPRs are the precursors of the normal dense cores, 
the fraction of CCPRs over normal cores would represent the duration or the age 
of the CCPR phase. Such a statistical estimation is valid only for the Taurus region 
because the number of both sample cores, CCPRs and normal ones, are available only in this region. 
If we simply assume a typical lifetime of starless NH$_{3}$ and/or CCS cores 
where the typical H$_{2}$ density is an order of 10$^{4}$~cm$^{-3}$ \citep{suzuki1992}, 
to be (2-10)$\times10^{5}$~yr \citep[e.g., Figure 2 in][]{ward-thompson2007}, 
the CCPR phase would last no longer than $\sim$(0.4-2)$\times10^{5}$~yr. 
The lifetime is determined from the fraction of CCPRs with respect 
to all the starless cores, 4/20, in our sample. 
This value is in good agreement with the ratio of the duration 
of CCPRs and other cores, $\sim4\times10^{5}$~yr and $\sim20\times10^{5}$~yr, 
respectively, estimated from a chemical model calculations and previous 
observational results \citep{ohishi1998}. 
Thus, our statistically estimated lifetime of the CCPR phase in the Taurus cores 
is consistent with that of theoretical prediction. 

If stars as well as cores have been forming steadily at a constant rate 
in both the Taurus and Ophiuchus regions with the same evolutionary timescale, 
we would expect to detect three CCPRs ($\sim$20\%) among 
15 starless cores in our Ophiuchus sample (Table \ref{tab-detection}). 
None detection of CCPRs in the Ophiuchus region would imply that there is no core 
younger than $\sim10^{5}$~yr in Ophiuchus, 
although we cannot completely rule out the possibility of a selection bias. 
In other words, dense core formation in Ophiuchus might have stopped 
since $\sim10^{5}$~yr ago which results in a true deficient of CCPRs in 
the Ophiuchus region. 
A similar picture is proposed by \citet{jorgensen2008} on the basis of 
the fact that fraction of embedded young stellar objects is relatively 
lower in the Ophiuchus region than in Perseus. 
However, there still remains sufficient amount of 
molecular gas traced by the $^{13}$CO and C$^{18}$O lines \citep{tachihara2000} 
which have potential to form more dense cores, 
and hence, it seems unlikely that the dense core formation has 
stopped over $\sim10^{5}$~yr. 

\subsubsection{Difference in Contraction Timescales of the Cores}

Another possible scenario for the systematic 
molecular abundance variations is that the difference in the 
contraction timescale between the Taurus cores and Ophiuchus cores. 
The contraction timescale would reflect the physical properties such 
as average density, magnetic field strength and/or 
turbulence in/around the cores \citep[e.g.,][]{aikawa2001, aikawa2003, aikawa2005}. 
According to the detailed theoretical model calculations, 
behaviors of molecular abundances and their distributions are predicted to 
depend strongly on the timescale of 
the gravitational contraction; the longer contraction time would result in 
the less abundances of carbon-chain molecules at the central part of the core 
\citep{aikawa2001, aikawa2003, aikawa2005}. 
Therefore, the systematically lower abundances 
of carbon-chain molecules in the Ophiuchus cores could be accounted for by 
the slower contraction than in the Taurus cores. 

Previous works have shown that the star-formation activity in the Ophiuchus region 
seems to be remarkably different from that in Taurus. 
There exist an active cluster-forming region in Ophiuchus triggered by a nearby Scorpius 
OB association, while only isolated star-formation can be seen in Taurus 
\citep[e.g.,][]{ward-thompson2007}. 
As a consequence, one would expect slower contraction timescale in the Taurus region than in 
Ophiuchus, which is exactly the opposite of what we are proposing in this section. 
However, active cluster-formation is localized only around "the main body" in 
the $\rho$-Ophiuchus molecular cloud or a dark cloud L1688, 
and rest of the region shows rather inactive star-formation 
\citep{tachihara2000}. 
In fact, overall star-formation efficiency in 
the Ophiuchus region, in particular Ophiuchus north region, is lower than in 
Taurus \citep{tachihara2000}. 
Based on the survey of the C$^{18}$O line in the Ophiuchus region, 
timescale of starless cores in the Ophiuchus region is estimated to be 
8$\times$10$^{5}$ yr, which is four times longer than that 
in Taurus, 2$\times$10$^{5}$ yr \citep{tachihara2000}. 
It is consistent with our hypothesis that 
the contraction timescale in Taurus is generally shorter than in Ophiuchus. 
This difference might be related to 
the difference in the line widths of the C$^{18}$O lines 
which reflect turbulence of the cores; 
mean line widths are 0.49, 0.9, and 0.7~km~s$^{-1}$ for the Taurus, $\rho$-Ophiuchus 
main body, and Ophiuchus north regions, respectively \citep{tachihara2000}. 

One might suppose that the difference in initial conditions 
can be an origin of chemical abundance variation. 
Due to higher initial densities in the $\rho$-Ophiuchus main body 
than in Taurus \citep{jijina1999, tachihara2000}, 
chemical evolution in the molecular cloud could occur prior to dense core formation in 
the $\rho$-Ophiuchus main body. In this case, cores 
in the $\rho$-Ophiuchus main body would be formed from already processed gas 
(e.g., higher NH$_{3}$/CCS ratios), while Taurus ones from more pristine material 
(lower NH$_{3}$/CCS ratio). This could produce the observed dichotomy 
without any need for a faster core contraction in Taurus. 
Higher initial density in the $\rho$-Ophiuchus main body region 
would results in faster gravitational contraction than in Taurus, 
giving a high star-formation rate. 
However, it is predicted that initial density of molecular cloud prior 
to the core formation ($<10^{4}$~cm$^{-3}$) would not much affect 
the molecular abundances in the dense core \citep{aikawa2003}. 
Therefore, we prefer the possible scenario of 
the faster contraction in the Taurus regions than in Ophiuchus 
rather than the difference in the initial densities. 

In our samples, there are five cores which are embedded in the $\rho$-Ophiuchus 
main body, L1681A, L1681B, L1690, L1696A, and L1696B (Table \ref{tab-summary}). 
Among them, only L1681B and L1696A were detected in the NH$_{3}$ line 
with normal NH$_{3}$/CCS ratios while other cores were not detected 
probably due to the position offset from real core center. 
Further complete survey of the Ophiuchus cores \citep[e.g., L1689B observed by][]{lee2003} 
is still an important issue to shed light on chemical and dynamical timescale 
of dense cores in this region. 

The difference in timescale of evolution of dense cores and 
young stellar objects is also proposed by \citet{jorgensen2008} based on the 
comparison of young stellar objects and dense cores in the Ophiuchus and 
Perseus regions. Thus, it is interesting to compare statistically physics and chemistry 
of dense cores and young stellar objects between Taurus and Ophiuchus regions. 

The difference in the contraction timescale is also reflected in the molecular 
distributions of CCS and NH$_{3}$ in the observed cores. 
In general, depletion of molecules 
at the central part of the core is more significant for the slower contraction case, 
and hence, the size of the central hole of the distribution of CCS becomes larger 
\citep{aikawa2001, aikawa2003, aikawa2005}. 
In contrast, cores with fast contraction show centrally condensed distribution 
of CCS without significant depletion \citep{aikawa2005}. 
For instance, a chemical and dynamical model with a faster contraction is 
preferable for 
L1521E, one of the CCPRs, \citep{aikawa2005} 
judging from its central density, chemical composition along with no signature of 
depletion, and slightly broader line width than in other prestellar cores. 
In our mapping observations, CCS distributions in two carbon-chain rich cores in the Aquila region, 
L492 and L483, are found to be centrally peaked structure. 
The contraction timescales of the L492 and L483 cores would have been short 
although they are dynamically evolved; 
L492 is in the dynamically collapsing 
\citep{lee2001, hirota2006} and L483 has already formed Class 0 protostar \citep{fuller1995}. 
They may be even shorter than those of the Taurus cores considering 
the possible difference in the chemical and dynamical evolutionary 
stages of the cores between the Aquila rift and the Taurus region, 
as suggested by \citet{hirota2006}. 

In order to clarify the difference in the evolutionary timescale and/or age between 
the Aquila region and the Taurus region, 
it is crucial to observe the distribution of CCS in L492 and L483, as well as the 
known CCPRs in the Taurus molecular cloud, 
with the higher resolution interferometers. 
Such observations would enable us to confirm whether there are really no central holes of CCS 
in these cores \citep[e.g.,][]{velusamy1995}. 
Although it is difficult to infer the timescale or the age of each core quantitatively, 
our results would provide some hints to understand the timescale of 
the formation and evolution of dark cloud cores. 

\subsection{Warm Carbon-Chain Chemistry}

Finally, we comment on Warm Carbon-Chain Chemistry (WCCC) recently proposed by 
\citet{sakai2008-l1527}. 
The WCCC is first identified in 
L1527 and is characterized by high abundances of long carbon-chain molecules 
such as C$_{n}$H ($n$=4,5,6) and C$_{n}$H$_{2}$ ($n$=3,4,6), 
and intense spectra of higher excitation lines of carbon-chain molecules. 
According to these criteria, L483 is considered to be a possible candidate of the WCCC 
source. When it is compared with L1527, the both 
are associated with the {\it{IRAS}} point source, and 
the NH$_{3}$/CCS ratios are relatively high due to high NH$_{3}$ column 
density. The abundances of HC$_{3}$N are remarkably high and are 
comparable to those of CCPRs such as 
L1495B, L1521B, L1521E, the cyanopolyyne peak of TMC-1, and L492. 
Furthermore, we stress that L483 and L1527 show stronger emission of 
longer carbon-chain molecule HC$_{5}$N than those in CCPRs. 
Except for relatively high abundance of CCS, chemistry in L483 seems to be 
quite similar to that in L1527. 
Observational studies of WCCC have just been 
started, and hence, the reason for such similarities and differences will be investigated 
in the near future \citep{sakai2008-l1527, sakai2009}. 
Interestingly, the regions where the WCCC sources 
are found, Taurus (L1527) and Aquila (L483), include 
CCPRs such as L1495B, L1521B, L1521E, the cyanopolyyne peak of TMC-1, and L492. 
On the other hand, no WCCC sources are found currently 
in the Ophiuchus and Perseus regions \citep{sakai2008-l1527, sakai2009}. 
Although the discussion is based on a limited number of samples at this 
moment, the WCCC might have a close connection to the CCPRs. 
Therefore, chemistry of carbon-chain molecules would be indicative of 
difference in the physical and chemical properties among 
molecular cloud complexes. 

\section{Summary}

In the present study, we carried out survey observations of 
CCS, HC$_{3}$N, and HC$_{5}$N toward 40 dark cloud cores in total 
using the 45~m radio telescope of the NRO. 
Based on the previous result that the spectra of carbon-chain molecules 
tend to be intense toward the cores where the NH$_{3}$ lines are 
weak \citep{suzuki1992}, 
we primarily observed the dense cores where the NH$_{3}$ lines 
are weak, not detected, or not observed previously. 
The main results of this paper are summarized as follows: 

\begin{enumerate}
\item We detected the CCS, HC$_{3}$N, and HC$_{5}$N lines for the 
17, 17, and 5 cores out of the 40, 40, and 35 cores observed, respectively. 
According to the previous survey of NH$_{3}$ conducted by 
\citet{benson1989}, the NH$_{3}$ lines are detected toward 24 cores 
out of our 40 cores. 
The detection rates of the CCS, HC$_{3}$N, and HC$_{5}$N lines 
are lower than those of previous survey \citep{suzuki1992}, 
although the spectra of carbon-chain molecules had been expected to be 
intense toward the cores with weak or no NH$_{3}$ lines. 
\item Among the 40 cores of our survey, we found remarkably 
intense spectra of CCS, HC$_{3}$N, and HC$_{5}$N toward a 
starless dense core L492 in the Aquila rift. We have carried out 
follow-up observations of L492 and found that this source is 
chemically less evolved dark cloud core like CCPRs. 
\item In addition, we found possible candidates for CCPRs, 
where the NH$_{3}$/CCS ratios are as low as other known 
CCPRs: L1517D, L530D, L1147, and L1172B. 
Further confirmation observations are needed for these sources. 
\item Together with the present results, we compiled available 
results of the survey of 
carbon-chain molecules and NH$_{3}$ to investigate the variation of 
the molecular abundances in the observed samples. 
We found that the abundance ratios of carbon-chain molecules and NH$_{3}$, 
in particular the NH$_{3}$/CCS ratio, which is a good indicator of chemical evolutionary 
stage of dark cloud cores, 
tend to be higher in the Ophiuchus cores than in the Taurus region. 
\item The systematic variation can be seen in the detection rates 
of carbon-chain molecules. For instance, the detection rates of 
the CCS, HC$_{3}$N, and HC$_{5}$N lines are significantly higher in the Taurus 
region than in the Ophiuchus region, although the total number of samples, 
fraction of cores associated with {\it{IRAS}} sources, and 
the detection rate of the NH$_{3}$ lines are similar to each other. 
\item These chemical abundance variation found in our study may 
suggest the region-to-region difference in the contraction timescale 
rather than the difference in the evolutionary stage alone, 
which could be indicative of some differences in the physical 
properties such as average density, magnetic field strength and/or 
turbulence in/around the cores. 
\end{enumerate}

It is important but difficult to compare evolutionary timescale by the 
statistical discussion because the present study is not based on the unbiased survey. 
Further complete and unbiased survey \citep[e.g.,][]{rathborne2008, rosolowsky2008}, 
would be needed to understand low-mass star-formation processes in dark cloud cores 
both from the chemical and dynamical point of views. Mapping observations of 
some tracers of chemical evolution such as carbon-chain molecules, 
NH$_{3}$, and deuterated molecules, and submillimeter continuum emission 
including polarization would also be important to reveal their dynamical 
and chemical evolutionary stage as proposed in several literatures 
\citep[e.g., ][]{lee2003, shirley2005, aikawa2005, hirota2006}. 

\acknowledgements

We are grateful to Miho Ikeda and all the staff of NRO 
for their assistance in observations. 
We are also grateful to Nami Sakai for useful discussions about WCCC. 
We thank the referee for useful suggestions. 
TH thanks to the Inoue Foundation for Science  
for the financial support (Research Aid of Inoue Foundation for Science). 
MO is supported by the Japan Society for the Promotion of Science 
(JSPS) Core-to-Core program. 
This study is partly supported by Grant-in-Aid from 
The Ministry of Education, Culture, Sports, Science and Technology of Japan 
(No. 14204013, 15071201, and 19024070). 
The 45~m radio telescope is operated by NRO, 
a branch of National Astronomical Observatory of Japan. 

{\it Facilities:} \facility{No:45m}.

{}

\newpage

\begin{deluxetable}{llllll}
\tabletypesize{\scriptsize}
\tablenum{1}
\tablewidth{0pt}
\tablecaption{Observed Lines 
\label{tab-obsedf}}
\tablehead{
\colhead{Molecule} & \colhead{Transition} &
  \colhead{$\nu$ (MHz)} & \colhead{$S_{ul}$\tablenotemark{b}} &
  \colhead{$\mu$ (Debye)} & 
  \colhead{Reference} 
}
\startdata
CCS & $J_{N}$=2$_{1}$-1$_{0}$ & 22344.033 & 1.98 & 2.81 & 1,2 \\
    & $J_{N}$=4$_{3}$-3$_{2}$ & 45379.033 & 3.97 & 2.81 & 1,2 \\
HC$_{3}$N & $J$=5-4 & 45490.316\tablenotemark{a} & 5.00 & 3.72 & 3 \\
HC$_{5}$N & $J$=17-16 & 45264.721 & 17.00 & 4.33 & 4 \\
NH$_{3}$ & (1,1;VH)\tablenotemark{c} & 23692.955 & \nodata & \nodata & \nodata \\
         & (1,1;H)\tablenotemark{c}  & 23693.895 & \nodata & \nodata & \nodata \\
         & (1,1;M)\tablenotemark{c}  & 23694.496 & 1.50 & 1.46 & 5 \\
         & (1,1;L)\tablenotemark{c}  & 23695.095 & \nodata & \nodata & \nodata \\
         & (1,1;VL)\tablenotemark{c} & 23696.037 & \nodata & \nodata & \nodata \\
\enddata
\tablenotetext{ a}{ Main hyperfine component}
\tablenotetext{ b}{ Intrinsic line strength}
\tablenotetext{ c}{Hyperfine components are blended. 
Each component is labeled as adopted by \citet{ungerechts1980}. }
\tablerefs{1: \citet{murakami1990}; 2: \citet{yamamoto1990}; 3: \citet{lafferty1978}; 
 4: \citet{alexander1976}; 5: \citet{cohen1974}}
\end{deluxetable}

\clearpage

\begin{deluxetable}{lccccl}
\tabletypesize{\scriptsize}
\tablenum{2}
\tablewidth{0pt}
\tablecaption{Observed Sources \label{tab-source}}
\tablehead{
\colhead{Source} & \colhead{$\alpha$(J2000)} &
 \colhead{$\delta$(J2000)}  & \colhead{{\it{IRAS}}} & \colhead{} & \colhead{}
\vspace{2mm} \\
\colhead{Name} & \colhead{(  h  m  s)} &
 \colhead{ (  $^{\circ}$ \hspace{0.3em} \arcmin 
   \hspace{0.3em} \arcsec)} & \colhead{Number} & \colhead{Reference} & \colhead{Date}
}
\startdata
L1495C  & 04 13 30.7 & $+$28 15 55 & 04108+2803  & 1 & 2002 Mar. \\
L1495D  & 04 14 18.2 & $+$28 15 52 &  \nodata    & 1 & 2002 Mar. \\
L1506   & 04 18 31.1 & $+$25 19 25 &  \nodata    & 1 & 2002 Mar. \\
L1400B  & 04 24 46.6 & $+$55 01 52 &  \nodata    & 1 & 2002 Mar. \\
L1400E  & 04 28 28.5 & $+$54 47 36 &  \nodata    & 1 & 2002 Mar. \\
L1400F  & 04 29 51.4 & $+$54 14 19 &  \nodata    & 1 & 2002 Mar. \\
L1551A  & 04 30 58.1 & $+$18 17 10 &  \nodata    & 1 & 2002 Mar. \\
L1551   & 04 31 30.0 & $+$18 12 30 &  04287+1806 & 1 & 2002 Mar. \\
L1445   & 04 32 07.0 & $+$46 37 23 &  \nodata    & 2 & 2002 Mar. \\
L1517D  & 04 55 48.1 & $+$30 38 46 &  \nodata    & 1 & 2004 Apr. \\
L1523   & 05 06 22.9 & $+$31 41 19 &  \nodata    & 1 & 2002 Mar. \\
L1778A  & 15 39 27.5 & $-$07 10 08 &  \nodata    & 1 & 2002 Mar. \\
L183(N) & 15 54 09.2 & $-$02 49 42 &  \nodata    & 3 & 2002 Mar. \\
L1721   & 16 14 28.2 & $-$18 54 44 &  \nodata    & 1 & 2002 Mar. \\
L1719B  & 16 22 12.4 & $-$19 38 41 &  \nodata    & 1 & 2002 Mar. \\
L1690   & 16 27 46.4 & $-$24 16 59 &  \nodata    & 1 & 2002 Mar. \\
L1709A  & 16 30 50.8 & $-$23 41 03 &  \nodata    & 1 & 2002 Mar. \\
L1709C  & 16 33 53.4 & $-$23 38 32 &  \nodata    & 1 & 2002 Mar. \\
L158    & 16 47 23.2 & $-$13 59 21 &  16445-1352 & 1 & 2002 Mar. \\
L191    & 16 47 29.3 & $-$12 28 38 &  \nodata    & 1 & 2002 Mar. \\
L204F   & 16 47 48.4 & $-$11 56 56 &  \nodata    & 1 & 2002 Mar. \\
B68     & 17 22 38.8 & $-$23 50 02 &  \nodata    & 1 & 2002 Mar. \\
L492    & 18 15 46.1 & $-$03 46 13 &  \nodata    & 2 & 2002 Mar.\tablenotemark{a} \\
L429-1  & 18 17 05.6 & $-$08 13 30 &  \nodata    & 2 & 2004 Apr. \\
L483    & 18 17 29.7 & $-$04 39 38 &  18148-0440 & 4 & 2004 Apr. \\
L530H   & 18 49 28.5 & $-$04 57 40 &  \nodata    & 1 & 2002 Mar. \\
L530D   & 18 49 57.3 & $-$04 49 49 &  \nodata    & 1 & 2004 Apr. \\
L1147   & 20 40 31.8 & $+$67 21 45 &  \nodata    & 1 & 2004 Apr. \\
L1155H  & 20 43 06.5 & $+$67 46 26 &  \nodata    & 1 & 2002 Mar. \\
L1155D  & 20 43 49.8 & $+$67 36 29 &  \nodata    & 1 & 2002 Mar. \\
L1172D  & 21 02 09.0 & $+$67 53 54 & 21017+6742  & 1 & 2002 Mar. \\
L1172B  & 21 03 32.0 & $+$68 11 58 & \nodata     & 1 & 2002 Mar. \\
\enddata
\tablenotetext{a}{Follow-up observations were carried out in 2003 February and 2004 March 
\citep{hirota2006}.}
\tablerefs{1: \citet{benson1989}; 2: \citet{lee2001}; 
3: \citet{ungerechts1980}; 4: \citet{benson1998}}
\end{deluxetable}

\clearpage

\begin{deluxetable}{lcccl}
\tabletypesize{\scriptsize}
\tablenum{3}
\tablewidth{0pt}
\tablecaption{Previously Observed Sources \label{tab-source-app}}
\tablehead{
\colhead{Source} & \colhead{$\alpha$(J2000)} &
 \colhead{$\delta$(J2000)}  & \colhead{{\it{IRAS}}} & \colhead{}
\vspace{2mm} \\
\colhead{Name\tablenotemark{a}} & \colhead{(  h  m  s)} &
 \colhead{ (  $^{\circ}$ \hspace{0.3em} \arcmin 
   \hspace{0.3em} \arcsec)} & \colhead{Number} & \colhead{Date}
}
\startdata
L1527   & 04 39 53.6 & $+$26 03 05 &  04368+2557 & 1998 Apr. \\	   
L1512   & 05 04 09.7 & $+$32 43 09 &  \nodata    & 1999 May. \\
L183(S) & 15 54 09.2 & $-$02 51 39 &  \nodata    & 1999 Feb. \\	   
L1696A  & 16 28 31.5 & $-$24 19 08 &  \nodata    & 1999 Feb. \\
L1689N  & 16 32 22.8 & $-$24 28 33 &  16293-2422 & 1999 Feb. \\
L1157   & 20 39 06.6 & $+$68 02 13 &  20386+6751 & 1999 Apr. \\
L1155C  & 20 43 30.0 & $+$67 52 42 &  \nodata    & 1999 Apr. \\
L1228   & 20 59 30.6 & $+$78 22 49 &  21004+7811 & 1999 Apr. \\
\enddata
\tablenotetext{a}{All the sources are listed in \citet{hirota2001}.}
\end{deluxetable}

\clearpage

\begin{deluxetable}{lllllll}
\tabletypesize{\scriptsize}
\tablenum{4}
\tablewidth{0pt}
\tablecaption{Gauss Fit Parameters for the Detected Sources
\label{tab-obsccs}}
\tablehead{
\colhead{Source Name} & 
  \colhead{($\Delta \alpha$ \arcsec, $\Delta \delta$ \arcsec)} & 
  \colhead{Line} & 
  \colhead{$T_{a}^{*}$ (K)} & \colhead{$v_{lsr}$ (km s$^{-1}$)} &
  \colhead{$\Delta v$ (km s$^{-1}$)} &  \colhead{$T_{rms}$ (K)} }
\startdata
L1495C  & (+200,-160) & CCS($J_{N}$=4$_{3}$-3$_{2}$) & 0.91(23)  & 6.37(8)  & 0.60(18)  &  0.11 \\
        &             & HC$_{3}$N($J$=5-4)   & 1.32(14)  & 6.38(4)  & 0.82(10)  &  0.10 \\
        &             & HC$_{5}$N($J$=17-16) & $<$0.33   & \nodata  & \nodata   &  0.11 \\  
L1400B  & (-80,+40)   & CCS($J_{N}$=4$_{3}$-3$_{2}$) & 0.61(20)  & 3.29(6)  & 0.31(12)  &  0.07 \\
        &             & CCS($J_{N}$=2$_{1}$-1$_{0}$) & 0.21(11)  & 3.43(14) & 0.52(32)  &  0.04 \\
        &             & HC$_{3}$N($J$=5-4)   & 0.74(13)  & 3.33(6)  & 0.63(13)  &  0.07 \\
        &             & HC$_{5}$N($J$=17-16) & $<$0.20   & \nodata  & \nodata   &  0.07 \\  
        &             & NH$_{3}$(1,1;M)      & 0.26(13)  & 3.41(28) & 1.09(65)  &  0.05 \\
L1527   & (0,0)       & CCS($J_{N}$=4$_{3}$-3$_{2}$)\tablenotemark{a}
                                             & 0.43(17)  & 5.85(10) & 0.49(23)  &  0.08  \\
        &             & HC$_{3}$N($J$=5-4)   & 2.34(15)  & 6.01(3)  & 0.84(7)   &  0.08 \\
        &             & HC$_{5}$N($J$=17-16) & 0.58(14)  & 5.90(7)  & 0.57(17)  &  0.09 \\
L1517D  & (+80,+80)   & CCS($J_{N}$=4$_{3}$-3$_{2}$) & 0.42(11)  & 5.90(5)  & 0.37(12)  &  0.05 \\
        &             & CCS($J_{N}$=2$_{1}$-1$_{0}$) & $<$0.16   & \nodata  & \nodata   &  0.05 \\
        &             & HC$_{3}$N($J$=5-4)   & $<$0.14   & \nodata  & \nodata   &  0.05 \\  
        &             & HC$_{5}$N($J$=17-16) & $<$0.14   & \nodata  & \nodata   &  0.05 \\
        &             & NH$_{3}$(1,1;M)      & $<$0.13   & \nodata  & \nodata   &  0.04 \\
L1512   & (-40,0)     & CCS($J_{N}$=4$_{3}$-3$_{2}$) & 0.88(14)  & 7.17(3)  & 0.33(6)   &  0.08 \\
        &             & HC$_{3}$N($J$=5-4)   & 2.01(61)  & 7.15(11) & 0.69(27)  &  0.07 \\
        &             & HC$_{5}$N($J$=17-16)\tablenotemark{b}
                                             & \nodata   & \nodata  & \nodata   & \nodata \\
L183(N) & (0,0)       & CCS($J_{N}$=4$_{3}$-3$_{2}$) & 0.58(43)  & 2.49(10) & 0.23(22)  &  0.11  \\
        &             & HC$_{3}$N($J$=5-4)   & 1.42(20)  & 2.45(5)  & 0.73(12)  &  0.10  \\
        &             & HC$_{5}$N($J$=17-16) & $<$0.27   & \nodata  & \nodata   &  0.09  \\
L183(S) & (0,0)       & CCS($J_{N}$=4$_{3}$-3$_{2}$)\tablenotemark{a} 
                                             & 0.40(19)  & 2.53(18) & 0.73(42)  &  0.11  \\
        &             & HC$_{3}$N($J$=5-4)   & 0.63(20)  & 2.41(10) & 0.62(23)  &  0.11  \\
        &             & HC$_{5}$N($J$=17-16)\tablenotemark{b} 
                                             & \nodata   & \nodata  & \nodata   & \nodata \\
L1689N  & (0,0)       & CCS($J_{N}$=4$_{3}$-3$_{2}$)\tablenotemark{a}
                                             & 0.42(16)  & 3.96(11) & 0.57(27)  &  0.08  \\
        &             & HC$_{3}$N($J$=5-4)   & 0.67(11)  & 4.02(9)  & 1.05(23)  &  0.08  \\
        &             & HC$_{5}$N($J$=17-16)\tablenotemark{b}
                                             & \nodata   & \nodata  & \nodata   & \nodata \\
B68     & (0,0)       & CCS($J_{N}$=4$_{3}$-3$_{2}$) & 0.41(24)  & 3.32(9)  & 0.30(21)  &  0.09 \\
        &             & HC$_{3}$N($J$=5-4)   & 1.65(17)  & 3.37(4)  & 0.79(10)  &  0.08 \\
        &             & HC$_{5}$N($J$=17-16) & 0.38(19)  & 3.34(10) & 0.38(23)  &  0.09 \\
L492    & (40,0)      & CCS($J_{N}$=$4_{3}$-$3_{2}$)\tablenotemark{c}        & 1.51(10) & 7.73(2) & 0.48(4)  & 0.07 \\
        &             & CCS($J_{N}$=$2_{1}$-$1_{0}$)\tablenotemark{c}        & 0.69(6)  & 7.78(3) & 0.80(8)  & 0.03 \\
        &             & HC$_{3}$N($J$=5-4)\tablenotemark{c}                  & 2.35(16) & 7.76(3) & 0.81(7)  & 0.09 \\
        &             & HC$_{5}$N($J$=17-16)\tablenotemark{c}                & 1.23(12) & 7.76(2) & 0.49(6)  & 0.06 \\
        & (0,0)       & NH$_{3}$(1,1;VH)\tablenotemark{c,d}                  & 0.19(7)  & 7.87(4) & 0.86(10) & 0.04 \\
        &             & NH$_{3}$(1,1;H)\tablenotemark{c,d}                   & 0.31(7)  & 7.87(4) & 0.86(10) & 0.04 \\
        &             & NH$_{3}$(1,1;M)\tablenotemark{c,d}                   & 0.74(8)  & 7.87(4) & 0.86(10) & 0.04 \\
        &             & NH$_{3}$(1,1;L)\tablenotemark{c,d}                   & 0.33(7)  & 7.87(4) & 0.86(10) & 0.04 \\
        &             & NH$_{3}$(1,1;VL)\tablenotemark{c,d}                  & 0.24(7)  & 7.87(4) & 0.86(10) & 0.04 \\
L429-1 & (0,-40)      & CCS($J_{N}$=4$_{3}$-3$_{2}$) & 0.76(10)  & 6.85(5)  & 0.66(11)  &  0.11 \\
       &              & HC$_{3}$N($J$=5-4)   & 0.78(16)  & 6.77(9)  & 0.88(21)  &  0.08 \\
       &              & HC$_{5}$N($J$=17-16) & $<$0.24   & \nodata  & \nodata   &  0.08 \\
       & (0,0)        & NH$_{3}$(1,1;VH)\tablenotemark{d}     & 0.47(12)  & 6.87(7)  & 1.43(17)  &  0.09 \\
       &              & NH$_{3}$(1,1;H)\tablenotemark{d}      & 0.59(12)  & 6.87(7)  & 1.43(17)  &  0.09 \\
       &              & NH$_{3}$(1,1;M)\tablenotemark{d}      & 1.05(13)  & 6.87(7)  & 1.43(17)  &  0.09 \\
       &              & NH$_{3}$(1,1;L)\tablenotemark{d}      & 0.46(12)  & 6.87(7)  & 1.43(17)  &  0.09 \\
       &              & NH$_{3}$(1,1;VL)\tablenotemark{d}     & 0.35(12)  & 6.87(7)  & 1.43(17)  &  0.09 \\
L483   & (0,0)       & CCS($J_{N}$=4$_{3}$-3$_{2}$)  & 0.84(25)  & 5.31(8)  & 0.49(17)  &  0.15 \\
       &             & CCS($J_{N}$=2$_{1}$-1$_{0}$)  & 0.36(16)  & 5.36(22) & 1.00(56)  &  0.06 \\
       &             & HC$_{3}$N($J$=5-4)    & 2.03(18)  & 5.35(4)  & 0.85(9)   &  0.12 \\
       &             & HC$_{5}$N($J$=17-16)  & 0.75(21)  & 5.23(6)  & 0.40(14)  &  0.12 \\
       &             & NH$_{3}$(1,1;VH)\tablenotemark{d}      & 1.16(13)  & 5.54(3)  & 1.06(6)   &  0.05 \\
       &             & NH$_{3}$(1,1;H)\tablenotemark{d}       & 1.18(13)  & 5.54(3)  & 1.06(6)   &  0.05 \\
       &             & NH$_{3}$(1,1;M)\tablenotemark{d}       & 2.16(14)  & 5.54(3)  & 1.06(6)   &  0.05 \\
       &             & NH$_{3}$(1,1;L)\tablenotemark{d}       & 1.35(13)  & 5.54(3)  & 1.06(6)   &  0.05 \\
       &             & NH$_{3}$(1,1;VL)\tablenotemark{d}      & 0.86(13)  & 5.54(3)  & 1.06(6)   &  0.05 \\
L530D   & (+80,+40)  & CCS($J_{N}$=4$_{3}$-3$_{2}$)  & 0.36(15)  & 3.26(10) & 0.47(24)  &  0.08 \\
        &            & CCS($J_{N}$=2$_{1}$-1$_{0}$)  & $<$0.11   & \nodata  & \nodata   &  0.04 \\
        &            & HC$_{3}$N($J$=5-4)    & $<$0.20   & \nodata  & \nodata   &  0.07 \\  
        &            & HC$_{5}$N($J$=17-16)  & $<$0.20   & \nodata  & \nodata   &  0.07 \\
        &            & NH$_{3}$(1,1;M)       & $<$0.08   & \nodata  & \nodata   &  0.03 \\
L1157   & (0,0)      & CCS($J_{N}$=4$_{3}$-3$_{2}$)\tablenotemark{a}
                                             & $<$0.13   & \nodata  & \nodata   &  0.04 \\
        &            & HC$_{3}$N($J$=5-4)    & 0.18(3)   &  2.71(9) & 0.96(6)   &  0.04 \\
        &            & HC$_{5}$N($J$=17-16)\tablenotemark{b}
                                             & \nodata   & \nodata  & \nodata   & \nodata \\
L1147   & (0,-80)    & CCS($J_{N}$=4$_{3}$-3$_{2}$)  & 0.54(18)  & 2.79(7)  & 0.40(17)  &  0.08 \\
        &            & CCS($J_{N}$=2$_{1}$-1$_{0}$)  & 0.25(08)  & 2.81(10) & 0.56(22)  &  0.03 \\
        &            & HC$_{3}$N($J$=5-4)    & 0.64(12)  & 2.80(6)  & 0.66(14)  &  0.06 \\
        &            & HC$_{5}$N($J$=17-16)  & $<$0.22   & \nodata  & \nodata   &  0.07 \\
        &            & NH$_{3}$(1,1;M)       & 0.08(8)   & 2.64(26) & 0.46(56)  &  0.02 \\
L1155C  & (0,0)       & CCS($J_{N}$=4$_{3}$-3$_{2}$)\tablenotemark{a} 
                                             & 0.34(13)  & 2.80(12) & 0.64(29)  &  0.05 \\
        &             & HC$_{3}$N($J$=5-4)   & 0.30(9)   & 2.84(12) & 0.81(28)  &  0.05 \\
        &             & HC$_{5}$N($J$=17-16)\tablenotemark{a} 
                                             & \nodata   & \nodata  & \nodata   & \nodata \\
L1228   & (0,0)       & CCS($J_{N}$=4$_{3}$-3$_{2}$)\tablenotemark{a}
                                             & $<$0.18   & \nodata  & \nodata   &  0.06 \\
        &             & HC$_{3}$N($J$=5-4)   & 0.44(16)  & -7.43(10) & 0.53(24) &  0.07 \\
        &             & HC$_{5}$N($J$=17-16)\tablenotemark{a}
                                             & \nodata   & \nodata  & \nodata   & \nodata \\
L1172D  & (+80,+40)  & CCS($J_{N}$=4$_{3}$-3$_{2}$)  & 0.20(9)   & 2.84(26) & 1.10(61)  &  0.07 \\
        &            & HC$_{3}$N($J$=5-4)    & 1.55(10)  & 2.83(3)  & 0.98(7)   &  0.06 \\
        &            & HC$_{5}$N($J$=17-16)  & 0.36(12)  & 2.84(12) & 0.73(28)  &  0.06 \\
L1172B  & (-120,-40) & CCS($J_{N}$=4$_{3}$-3$_{2}$)  & 0.25(11)  & 2.42(17) & 0.71(39)  &  0.07 \\
        &            & CCS($J_{N}$=2$_{1}$-1$_{0}$)  & 0.21(08)  & 2.26(18) & 0.90(42)  &  0.04 \\
        &            & HC$_{3}$N($J$=5-4)    & 0.51(12)  & 2.38(9)  & 0.83(22)  &  0.06 \\
        &            & HC$_{5}$N($J$=17-16)  & $<$0.20   & \nodata  & \nodata   &  0.07 \\
        &            & NH$_{3}$(1,1;M)       & 0.20(5)   & 2.48(26) & 1.92(61)  &  0.04 \\
        & (-40,-40)  & CCS($J_{N}$=4$_{3}$-3$_{2}$)  & 0.38(16)  & 2.36(11) & 0.50(26)  &  0.07 \\
        &            & CCS($J_{N}$=2$_{1}$-1$_{0}$)  & 0.17(8)   & 2.34(21) & 0.88(48)  &  0.03 \\
        &            & HC$_{3}$N($J$=5-4)    & 0.77(11)  & 2.35(06) & 0.76(13)  &  0.06 \\
        &            & HC$_{5}$N($J$=17-16)  & $<$0.19   & \nodata  & \nodata   &  0.06 \\
        &            & NH$_{3}$(1,1;M)       & $<$0.12   & \nodata  & \nodata   &  0.04 \\
        & (+40,+40)  & CCS($J_{N}$=4$_{3}$-3$_{2}$)  & 0.45(13)  & 2.69(9)  & 0.58(20)  &  0.06 \\
        &            & CCS($J_{N}$=2$_{1}$-1$_{0}$)  & 0.19(10)  & 2.75(14) & 0.49(31)  &  0.04 \\
        &            & HC$_{3}$N($J$=5-4)    & 0.45(9)   & 2.69(9)  & 0.92(22)  &  0.06 \\
        &            & HC$_{5}$N($J$=17-16)  & $<$0.18   & \nodata  & \nodata   &  0.06 \\
        &            & NH$_{3}$(1,1;M)       & $<$0.12   & \nodata  & \nodata   &  0.04 \\
\enddata
\tablecomments{The numbers in parenthesis represent 
three times the standard deviation in the Gaussian fit.}
\tablenotetext{a}{\citet{hirota2001}}
\tablenotetext{b}{Not observed. }
\tablenotetext{c}{\citet{hirota2006}}
\tablenotetext{d}{The LSR velocity $v_{lsr}$ and line width $\Delta v$ are 
common for all the hyperfine components.} 
\end{deluxetable}

\clearpage

\begin{deluxetable}{llcccccc}
\tabletypesize{\scriptsize}
\tablenum{5}
\tablewidth{0pt}
\tablecaption{Column Densities of CCS, HC$_{3}$N, and HC$_{5}$N Determined 
by the LTE Model \label{tab-collte}}
\tablehead{
\colhead{Source} &  & \colhead{$N$[CCS]} & \colhead{$N$[HC$_{3}$N]} & 
 \colhead{$N$[HC$_{5}$N]} & \colhead{$N$[NH$_{3}$]} 
 \vspace{2mm} \\
\colhead{Name} & 
\colhead{($\Delta \alpha$ \arcsec, $\Delta \delta$ \arcsec)} & 
\colhead{(10$^{12}$~cm$^{-2}$)} & 
\colhead{(10$^{12}$~cm$^{-2}$)} & \colhead{(10$^{12}$~cm$^{-2}$)} & 
\colhead{(10$^{14}$~cm$^{-2}$)} & \colhead{NH$_{3}$/CCS} & \colhead{Reference} 
}
\startdata
L1495C 	&	(+200,-160)	&	17.2	&	13.8	&	$<$9.0 	&	4.0    	& 23      &	1,1,1,2	\\
L1495D 	&	(0,0)  	&	$<$2.2	&	$<$1.4	&	$<$4.9 	&	$<$1.0 	& \nodata &	1,1,1,2	\\
L1506  	&	(0,0)  	&	$<$2.8	&	$<$1.7	&	$<$6.2 	&	$<$0.5 	& \nodata &	1,1,1,2	\\
L1400B 	&	(-80,+40)	&	5.1	&	5.0	&	$<$5.3 	&	0.62   	& 12      &	1,1,1,1	\\
L1400E 	&	(0,0)  	&	$<$2.6	&	$<$1.6	&	$<$5.6 	&	$<$0.82	& \nodata &	1,1,1,2	\\
L1400F 	&	(0,0)  	&	$<$2.4	&	$<$1.6	&	$<$5.1 	&	0.84   	& $>$35   &	1,1,1,2	\\
L1551A 	&	(0,0)  	&	$<$2.6	&	$<$1.5	&	$<$5.5 	&	$<$0.98	& \nodata &	1,1,1,2	\\
L1551  	&	(0,0)  	&	$<$2.3	&	$<$1.4	&	$<$5.0 	&	$<$0.72	& \nodata &	1,1,1,2	\\
L1445  	&	(0,0)  	&	$<$2.3	&	$<$1.6	&	$<$5.3 	&	\nodata	& \nodata &	1,1,1,*	\\
L1527  	&	(0,0)  	&	5.1	&	47	&	18.9   	&	5.0    	& 98      &	3,1,1,2	\\
L1517D 	&	(+80,+80)	&	3.7	&	$<$1.1	&	$<$3.7 	&	$<$0.13	& $<${\bf{3.5}}  &	1,1,1,1	\\
L1512  	&	(-40,0)	&	9.0	&	25  	&	$<$2.1 	&	7.0    	& 78      &	1,1,4,4	\\
L1523  	&	(0,0)  	&	$<$2.2	&	$<$1.4	&	$<$4.8 	&	$<$1.06	& \nodata &	1,1,1,2	\\
L1778A 	&	(0,0)  	&	$<$3.0	&	$<$1.6	&	$<$6.3 	&	$<$1.03	& \nodata &	1,1,1,2	\\
L183(N)	&	(0,0)  	&	3.6	&	13.7	&	$<$7.3 	&	16      	& 444     &	1,1,1,5	\\
L183(S)	&	(0,0)  	&	7.2	&	4.1	&	\nodata	&	32     	& 444     &	3,1,*,5	\\
L1721  	&	(0,0)  	&	$<$3.1	&	$<$1.8	&	$<$7.0 	&	$<$0.81	& \nodata &	1,1,1,2	\\
L1719B 	&	(0,0)  	&	$<$3.3	&	$<$1.8	&	$<$7.0 	&	$<$0.87	& \nodata &	1,1,1,2	\\
L1690  	&	(0,0)  	&	$<$3.1	&	$<$1.8	&	$<$7.0 	&	$<$0.72	& \nodata &	1,1,1,2	\\
L1696A 	&	(0,0)  	&	$<$2.7	&	$<$2.1	&	$<$2.0 	&	5.0    	& $>$185  &	3,1,6,2	\\
L1709A 	&	(0,0)  	&	$<$3.3	&	$<$1.8	&	$<$7.2 	&	0.99    	& $>$30   &	1,1,1,2	\\
L1689N 	&	(0,0)  	&	5.8	&	7.5	&	\nodata	&	20     	& 345     &	3,1,*,7	\\
L1709C 	&	(0,0)  	&	$<$3.3	&	$<$1.7	&	$<$6.9 	&	1.14    	& $>$35   &	1,1,1,2	\\
L158   	&	(0,0)  	&	$<$2.8	&	$<$1.7	&	$<$7.0 	&	4.0    	& $>$143  &	1,1,1,2	\\
L191   	&	(0,0)  	&	$<$2.8	&	$<$1.7	&	$<$6.4 	&	0.71    	& $>$25   &	1,1,1,2	\\
L204F  	&	(0,0)  	&	$<$2.9	&	$<$1.7	&	$<$6.1 	&	1.43    	& $>$49   &	1,1,1,2	\\
B68    	&	(0,0)  	&	3.0	&	18.9	&	7.8    	&	2.5    	& 83      &	1,1,1,2	\\
L492   	&	(40,0)\tablenotemark{a}  	&	53	&	172	&	41     	&	3.4    	& {\bf{6.4}}     &	8,8,8,8	\\
L429-1  	&	(0,-40)\tablenotemark{a}	&	14.2	&	7.5	&	$<$6.4 	&	10.4    	& 73      &	1,1,1,1	\\
L483   	&	(0,0)  	&	12.3	&	31	&	17.6    	&	14.9   	& 121     &	1,1,1,1	\\
L530H   	&	(0,0)  	&	$<$3.8	&	$<$2.0	&	$<$8.4  	&	$<$0.65	& \nodata &	1,1,1,2	\\
L530D   	&	(+80,+40)	&	4.1	&	$<$1.5	&	$<$5.4  	&	$<$0.09	& $<${\bf{2.2}}  &	1,1,1,1	\\
L1157   	&	(0,0)  	&	$<$2.3	&	1.64	&	\nodata	&	5.2    	& $>$226  &	3,1,*,9	\\
L1147  	&	(0,-80)	&	5.5	&	4.5	&	$<$5.9 	&	0.08    	& 
  {\bf{1.5}}     &	1,1,1,1,	\\
L1155H 	&	(0,0)  	&	$<$2.4	&	$<$1.6	&	$<$5.4 	&	$<$0.82	& \nodata &	1,1,1,2	\\
L1155C 	&	(0,0)  	&	5.2	&	2.4	&	\nodata	&	5.0    	& 96      &	3,1,*,2	\\
L1155D 	&	(0,0)  	&	$<$2.8	&	$<$1.6	&	$<$5.7 	&	$<$0.34	& \nodata &	1,1,1,2	\\
L1228  	&	(0,0)  	&	$<$2.4	&	2.3	&	\nodata	&	1.74    	& $>$73   &	3,1,*,2	\\
L1172D 	&	(+80,+40)	&	4.9	&	21 	&	14.4   	&	7.9    	& 161     &	1,1,1,2	\\
L1172B  	&	(-120,-40)	&	4.0	&	4.4	&	$<$5.2 	&	0.81    	& 20      &	1,1,1,1	\\
       	&	(-40,-40)	&	4.6	&	6.4	&	$<$5.0 	&	$<$0.13	& $<${\bf{2.8}}  &	1,1,1,1	\\
       	&	(+40,+40)	&	6.4	&	4.2	&	$<$4.8 	&	$<$0.13	& $<${\bf{2.0}}  &	1,1,1,1	\\
\enddata
\tablecomments{Bold letters indicate the NH$_{3}$/CCS ratios lower than 10, which is 
a criterion for CCPRs. }
\tablenotetext{a}{Position for the column density of NH$_{3}$ is (0,0).}
\tablerefs{1: Present study; 2: \citet{benson1989}; 3: \citet{hirota2001}; 4: \citet{suzuki1992}; 
5: \citet{ungerechts1980}; 6: \citet{benson1983}; 7: \citet{mundy1990}; 8: \citet{hirota2006}; 
9: \citet{bachiller1993}}
\end{deluxetable}

\begin{deluxetable}{lccccccccccccccc}
\rotate
\tabletypesize{\tiny}
\tablenum{6}
\tablewidth{0pt}
\tablecaption{Detection Rate for CCS, HC$_{3}$N, HC$_{5}$N, 
   and NH$_{3}$ \label{tab-detection}}
\tablehead{
   & & &
 \multicolumn{2}{c}{CCS} &  & 
   \multicolumn{2}{c}{HC$_{3}$N} &  & 
   \multicolumn{2}{c}{HC$_{5}$N} &  & 
   \multicolumn{2}{c}{NH$_{3}$} & \\
& & \colhead{Star-forming} & & & & & & & &  \\
\cline{4-5}   \cline{7-8}   \cline{10-11}  \cline{13-14} 
\colhead{Sample} & \colhead{Total} & \colhead{cores} & 
   \colhead{Detect/Observed} & \colhead{Ratio} & &
   \colhead{Detect/Observed} & \colhead{Ratio} & &
   \colhead{Detect/Observed} & \colhead{Ratio} & &
   \colhead{Detect/Observed} & \colhead{Ratio} & $\frac{\mbox{Num. of CCS cores}}{\mbox{Num. of NH}_{3}\mbox{ cores}}$
}
\startdata
\citet{suzuki1992}\tablenotemark{a} & 49 & 22 & 27/49 & 0.55 & & 33/49 & 0.67 & & 14/49 & 0.29 & & 39/49 & 0.80 & 0.69 \\
Present study                       & 40 &  9 & 17/40 & 0.43 & & 17/40 & 0.43 & &  5/35 & 0.14 & & 24/39 & 0.62 & 0.71 \\
Compiled ensemble                   & 90 & 34 & 48/90 & 0.53 & & 46/83 & 0.55 & & 18/80 & 0.23 & & 66/89 & 0.74 & 0.73 \\
\quad Taurus                        & 29 &  9 & 18/29 & 0.62 & & 19/29 & 0.66 & & 12/29 & 0.41 & & 20/29 & 0.69 & 0.90 \\
\quad Ophiuchus                     & 24 &  9 &  6/24 & 0.25 & &  5/21 & 0.24 & &  0/22 & 0.00 & & 18/24 & 0.75 & 0.33 \\
\quad other than Taurus             & 61 & 25 & 30/61 & 0.49 & & 27/54 & 0.50 & &  6/51 & 0.12 & & 37/60 & 0.62 & 0.81 \\ 
\hline
Pipe Nebula\tablenotemark{b} & 46 &  1 & 13/46 & 0.28 & & \nodata & \nodata & & 4/46 & 0.09 & & 29/46 & 0.63 & 0.45 \\
Perseus\tablenotemark{c}    & 193 & $>$44 & 96/193 & 0.51 & & \nodata & \nodata & & \nodata & \nodata & & 162/193 & 0.84 & 0.59 \\
\enddata
\tablenotetext{a}{Two sources in the Orion region, L1641N and NGC2071N are included, which are not listed in Table \ref{tab-summary}.} 
\tablenotetext{b}{\citet{rathborne2008}. We identified the Pipe core 12 as the only the site of star-formation \citep{brooke2007}.} 
\tablenotetext{c}{\citet{rosolowsky2008}. We identified the star-forming cores according to \citet{enoch2006}, which are labeled as "Bolocam" sources in 
\citet{rosolowsky2008}. We tentatively regard the sources other than "Bolocam" in \citet{rosolowsky2008} 
as the starless cores. Therefore, the number of star-forming cores in the Perseus gives a lower limit. }
\end{deluxetable}

\clearpage

\clearpage

\begin{figure*}[htb]
\includegraphics[width=5cm]{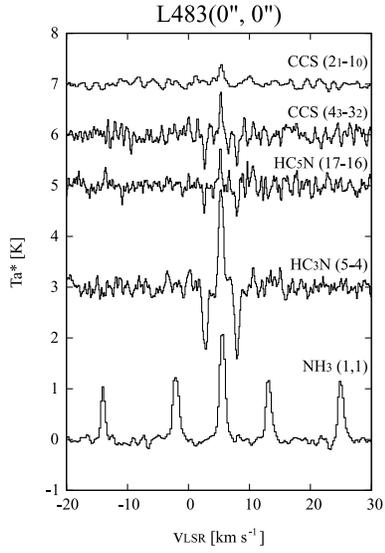}
\caption{Sample spectra of the observed lines in L483. 
The apparent features below the baseline of the CCS($J_{N}$=$4_{3}$-$3_{2}$), 
HC$_{3}$N($J$=5-4), and HC$_{5}$N($J$=17-16) spectra are artifacts 
of the frequency-switching technique. All of the five hyperfine components of the 
NH$_{3}$(1, 1) line are shown. 
\label{fig-l483}  }
\end{figure*}

\clearpage

\begin{figure*}[htb]
\includegraphics[width=15cm]{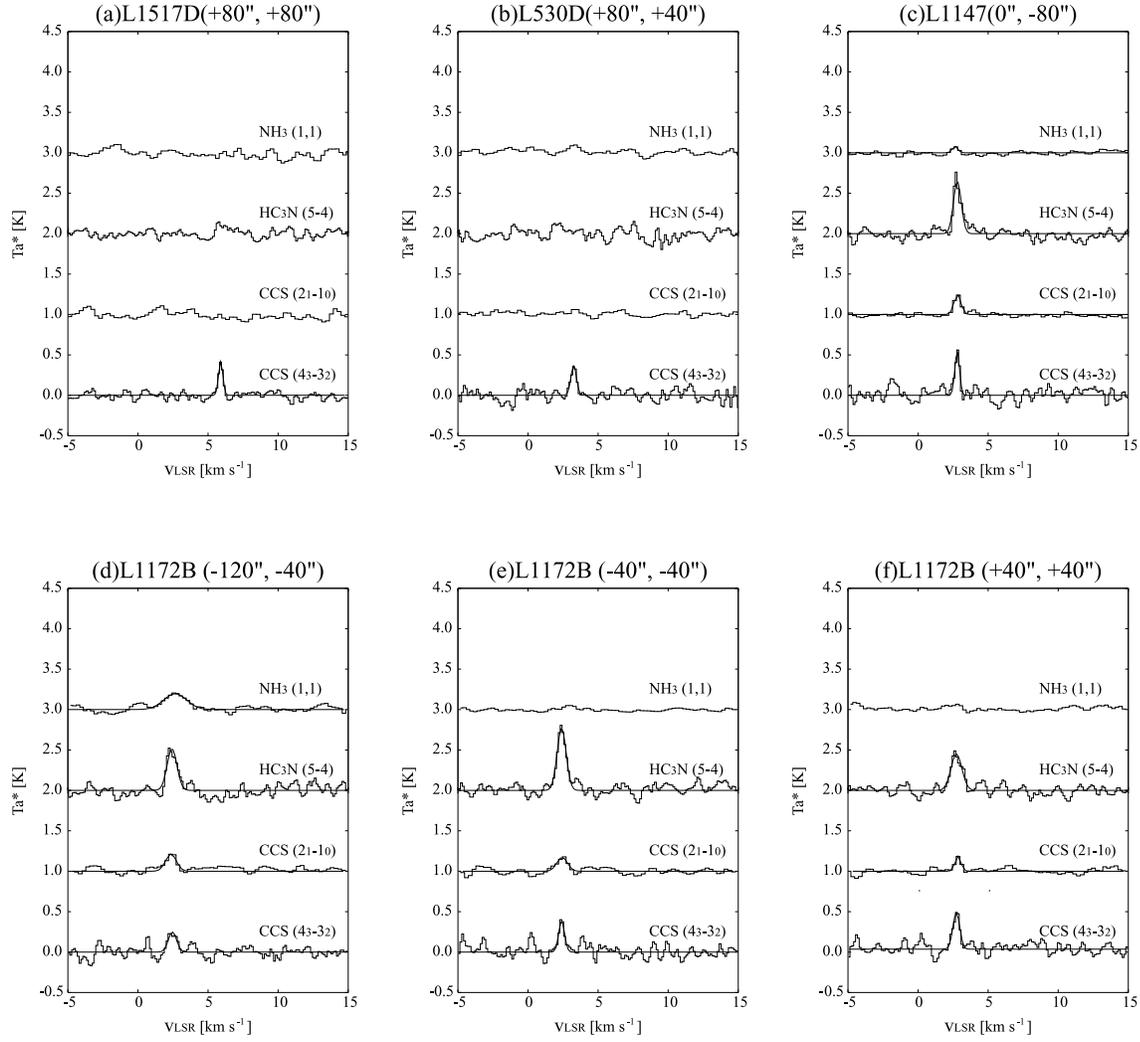}
\caption{Sample spectra of the observed lines in L1517D, L530D, L1147, 
and L1172B. Results of Gaussian fitting are also shown. 
\label{fig-sp}  }
\end{figure*}

\clearpage

\begin{figure*}[htb]
\includegraphics[width=15cm]{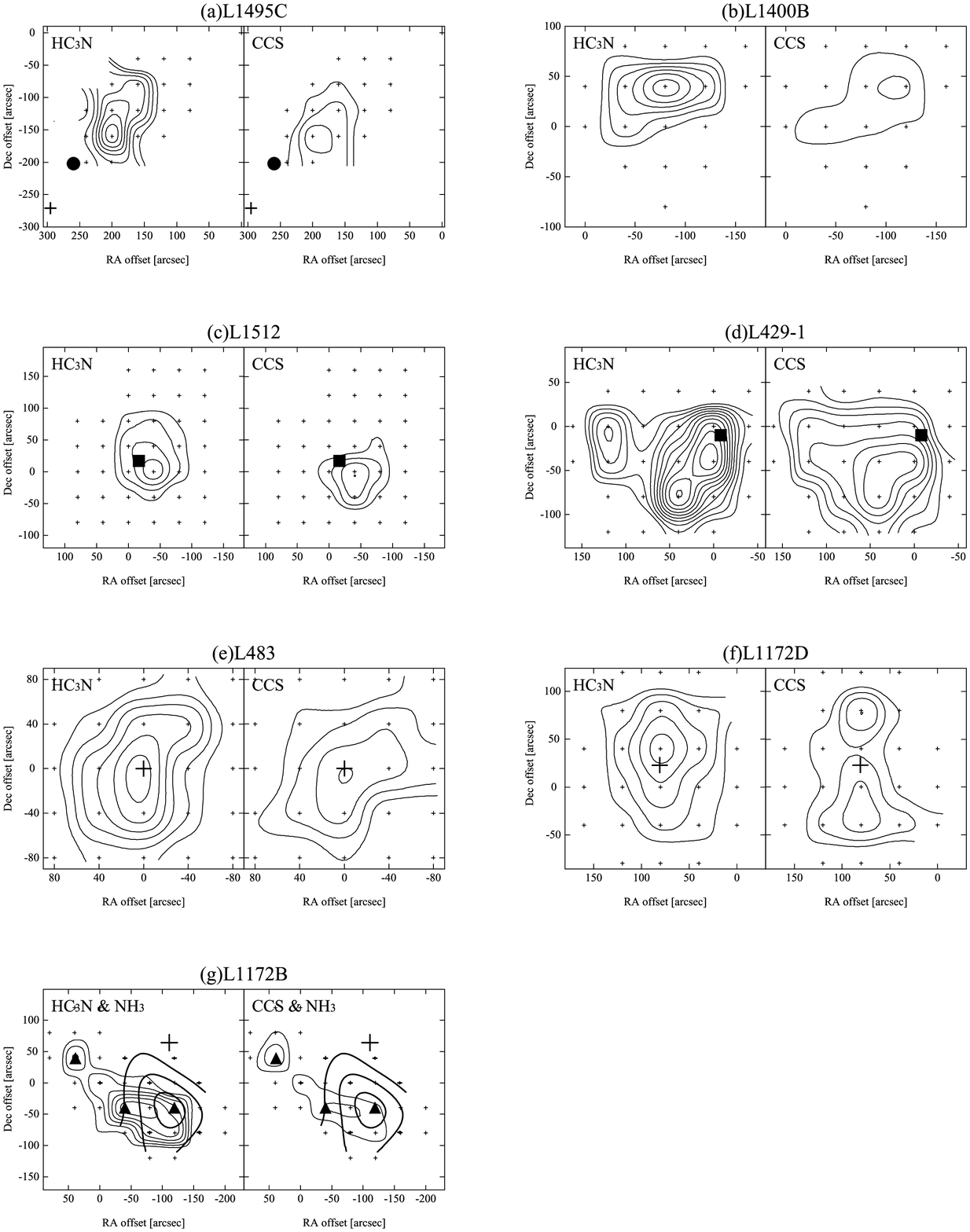}
\end{figure*}
\figcaption[]{Integrated intensity maps of HC$_{3}$N and CCS. 
(a) L1495C. A cross indicates the {\it{IRAS}} source ({\it{IRAS}} 04018+2803) 
and a filled circle represents the position of the NH$_{3}$ peak 
of L1495N \citep{benson1989}. 
{\it{Left:}} HC$_{3}$N($J$=5-4).  The velocity range of integration 
is 5.7-7.1 km s$^{-1}$. The interval of the contours 
is 0.1 K km s$^{-1}$ and the lowest one is 0.3 K km s$^{-1}$. 
{\it{Right:}} CCS($J_{N}$=$4_{3}$-$3_{2}$). The velocity range of integration 
is 5.7-6.9 km s$^{-1}$. The interval of the contours 
is 0.1 K km s$^{-1}$ and the lowest one is 0.3 K km s$^{-1}$. 
(b) L1400B. 
{\it{Left:}} HC$_{3}$N($J$=5-4).  The velocity range of integration 
is 2.8-4.0 km s$^{-1}$. The interval of the contours 
is 0.05 K km s$^{-1}$ and the lowest one is 0.15 K km s$^{-1}$. 
{\it{Right:}} CCS($J_{N}$=$4_{3}$-$3_{2}$). The velocity range of integration 
is 2.8-3.7 km s$^{-1}$. The interval of the contours 
is 0.05 K km s$^{-1}$ and the lowest one is 0.15 K km s$^{-1}$. 
(c) L1512. A filled square represents the peak position of the dust continuum 
emission \citep{kirk2005}.
{\it{Left:}} HC$_{3}$N($J$=5-4).  The velocity range of integration 
is 6.5-7.8 km s$^{-1}$. The interval of the contours 
is 0.2 K km s$^{-1}$ and the lowest one is 0.6 K km s$^{-1}$. 
{\it{Right:}} CCS($J_{N}$=$4_{3}$-$3_{2}$). The velocity range of integration 
is 6.6-7.7 km s$^{-1}$. The interval of the contours 
is 0.05 K km s$^{-1}$ and the lowest one is 0.15 K km s$^{-1}$. 
(d) L429-1. 
A filled square represents the peak position of the dust continuum 
emission \citep{crapsi2005}.
{\it{Left:}} HC$_{3}$N($J$=5-4).  The velocity range of integration 
is 6.0-7.7~km~s$^{-1}$. The interval of the contours 
is 0.05 K km s$^{-1}$ and the lowest one is 0.15 K km s$^{-1}$. 
{\it{Right:}} CCS($J_{N}$=$4_{3}$-$3_{2}$). The velocity range of integration 
is 6.3-7.5~km~s$^{-1}$. The interval of the contours 
is 0.05 K km s$^{-1}$ and the lowest one is 0.15 K km s$^{-1}$. 
(e) L483. 
A cross indicates the {\it{IRAS}} source ({\it{IRAS}} 18148-0440). 
{\it{Left:}} HC$_{3}$N($J$=5-4).  The velocity range of integration 
is 4.5-6.5~km~s$^{-1}$. The interval of the contours 
is 0.2 K km s$^{-1}$ and the lowest one is 0.6 K km s$^{-1}$. 
{\it{Right:}} CCS($J_{N}$=$4_{3}$-$3_{2}$). The velocity range of integration 
is 4.8-6.0~km s$^{-1}$. The interval of the contours 
is 0.07 K km s$^{-1}$ and the lowest one is 0.21 K km s$^{-1}$. 
(f) L1172D. 
A cross indicates the {\it{IRAS}} source ({\it{IRAS}} 21017+6742). 
{\it{Left:}} HC$_{3}$N($J$=5-4).  The velocity range of integration 
is 2.0-4.0 km s$^{-1}$. The interval of the contours 
is 0.2 K km s$^{-1}$ and the lowest one is 0.6 K km s$^{-1}$. 
{\it{Right:}} CCS($J_{N}$=$4_{3}$-$3_{2}$). The velocity range of integration 
is 2.2-3.5 km s$^{-1}$. The interval of the contours 
is 0.05 K km s$^{-1}$ and the lowest one is 0.15 K km s$^{-1}$. 
(g) L1172B. 
A cross indicates the {\it{IRAS}} source ({\it{IRAS}} 21025+6801). 
The three positions toward which the molecular abundances are obtained, 
(-120\arcsec,-40\arcsec), (-40\arcsec,-40\arcsec), and (+40\arcsec,+40\arcsec), 
are indicated by filled triangles. 
Bold contours represent the integrated intensity map of 
the NH$_{3}$ (1,1) lines with their interval and the lowest level 
are 0.044~K~km~s$^{-^1}$ and 0.132~K~km~s$^{-^1}$, respectively. 
{\it{Left:}} HC$_{3}$N($J$=5-4).  The velocity range of integration 
is 1.6-3.6 km s$^{-1}$. The interval of the contours 
is 0.07 K km s$^{-1}$ and the lowest one is 0.21 K km s$^{-1}$. 
{\it{Right:}} CCS($J_{N}$=$4_{3}$-$3_{2}$). The velocity range of integration 
is 1.8-3.3 km s$^{-1}$. The interval of the contours 
is 0.05 K km s$^{-1}$ and the lowest one is 0.15 K km s$^{-1}$. 
\label{fig-maps}  }

\clearpage

\begin{figure*}[htb]
\includegraphics[width=15cm]{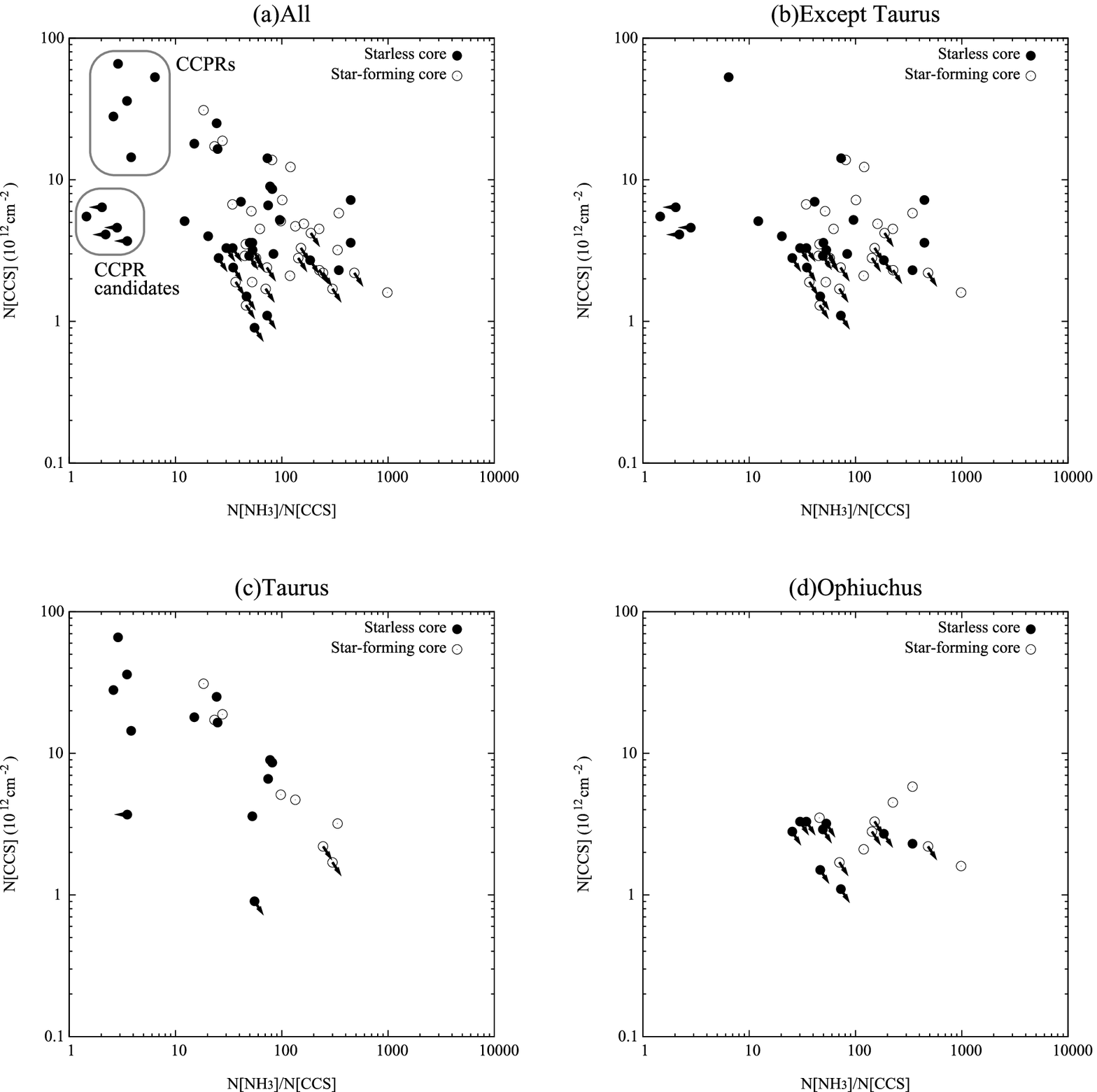}
\caption{Relationships between the NH$_{3}$/CCS ratios and 
the column densities of CCS. The cores where either CCS or NH$_{3}$ are detected are plotted. 
Filled and open symbols represent the cores without {\it{IRAS}} sources (starless cores) 
and the cores with {\it{IRAS}} sources (star-forming cores), respectively. 
Arrows represent the upper or lower limit of the values. 
(a) All samples listed in Table \ref{tab-summary}. 
(b) Same as (a) but except for Taurus cores. (c) Taurus cores. (d) Ophiuchus cores. 
\label{fig-nh3ccs}  }
\end{figure*}

\begin{figure*}[htb]
\includegraphics[width=15cm]{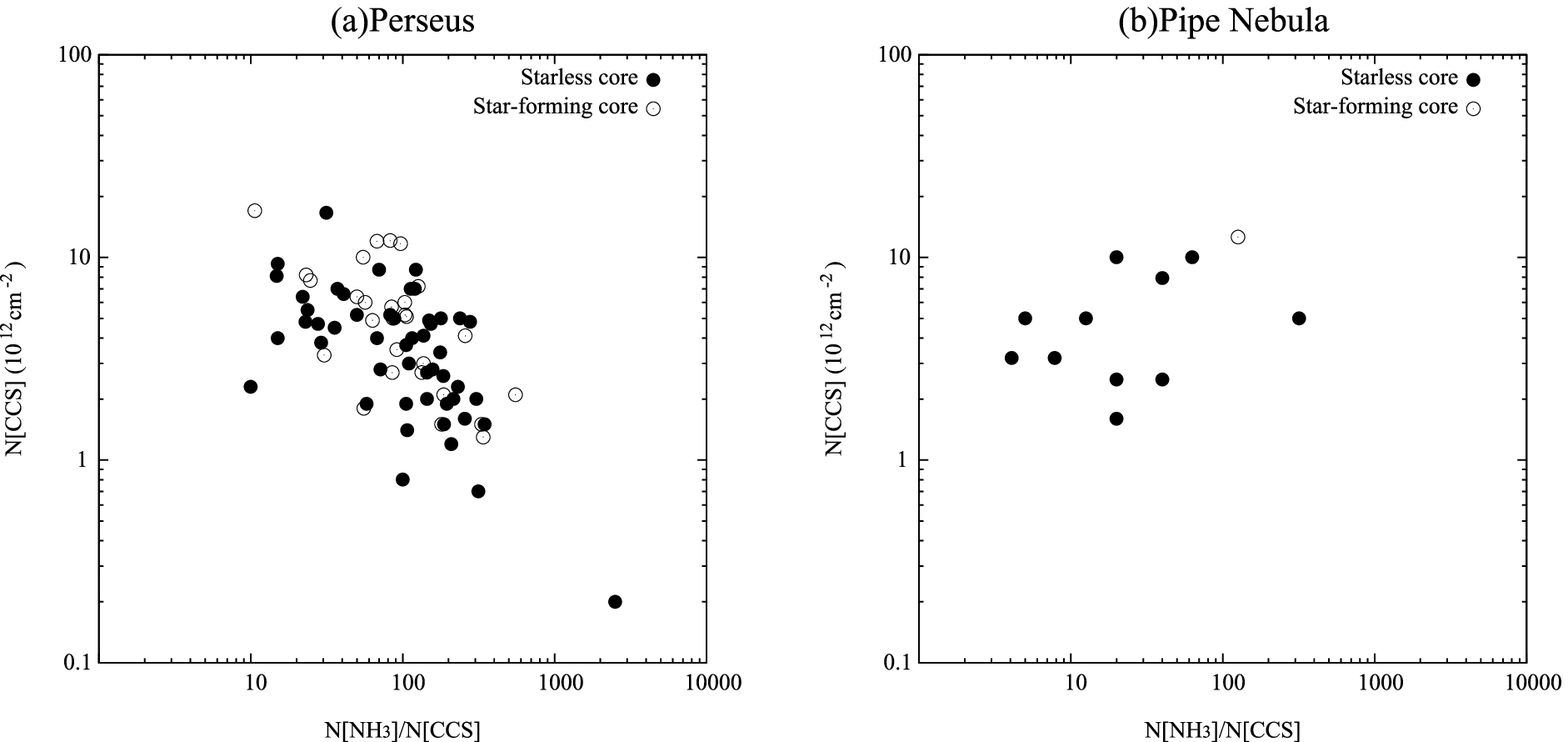}
\caption{Relationships between the NH$_{3}$/CCS ratios and 
the column densities of CCS. (a) Perseus cores. 
The star-forming cores are identified according to \citet{enoch2006}, 
which are labeled as "Bolocam" sources in \citet{rosolowsky2008}. 
We tentatively regard the sources other than "Bolocam" 
in \citet{rosolowsky2008} as the starless cores. Only the cores where 
both CCS and NH$_{3}$ are detected are plotted. 
(b) Pipe cores. In this panel, we indicated the Pipe core 12 as the only the 
site of star-formation \citep{brooke2007}. Only the cores where both 
CCS and NH$_{3}$ are detected are shown. 
\label{fig-nh3ccs-p}  }
\end{figure*}

\clearpage
\appendix
\section{Individual Sources}
\label{sec-individual}

In this appendix, we will describe the individual sources where at least one of 
the observed lines were detected. 

\subsection{L1495C}

This core, taken from \citet{benson1989}, 
is a starless core located in the Taurus molecular cloud at a distance of 137~pc 
\citep{torres2007}. 
We first detected the CCS and HC$_{3}$N lines at the (+80\arcsec, -80\arcsec) position 
in right ascension and declination, respectively, with respect to the reference 
position of L1495C. 
Then we extended our maps to about 4\arcmin \ southeast from the reference position, 
and found that the 
peak positions of CCS and HC$_{3}$N are located within the NH$_{3}$ core of L1495N 
observed by \citet{benson1989}. 
Its position offset is (-60\arcsec,+42\arcsec) with respect to the reference 
position of L1495N. Therefore, the CCS and HC$_{3}$N cores 
detected in the present study should be referred to L1495N rather than L1495C. 
The integrated intensity maps for CCS and HC$_{3}$N 
are shown in Figure \ref{fig-maps}(a). The {\it{IRAS}} source and 
the NH$_{3}$ core are located at the southeast edge of our map. 
Because the CCS and HC$_{3}$N peaks correspond to the northwestern edge 
of the NH$_{3}$ core of L1495N \citep{benson1989}, 
the carbon-chain molecules seem to be depleted at the NH$_{3}$ peak of L1495N. 

\subsection{L1400B}

This core is one of the starless cores listed in \citet{benson1989}, 
whose distance is 170~pc \citep{hilton1995}. 
We detected the CCS and HC$_{3}$N lines toward the 
(-80\arcsec, +40\arcsec) northwest from the reference position. 
The integrated intensity maps for CCS and HC$_{3}$N are shown in 
Figure \ref{fig-maps}(b). The peak position of CCS may be 
shifted from that of HC$_{3}$N, although it is unclear due to 
the low signal-to-noise ratio and the coarse sampling. 
We also detected the NH$_{3}$ lines toward the (-80\arcsec, +40\arcsec) position. 
Because the NH$_{3}$ lines were not detected by \citet{benson1989}, 
this is the first detection of the NH$_{3}$ lines toward L1400B. 
The line parameters are listed in Table \ref{tab-obsccs}. 

\subsection{L1527}

A dense core associated with L1527 in the Taurus molecular cloud 
is a host of Class 0 protostar {\it{IRAS}}~04368+2557, which is listed in \citet{benson1989}. 
The CCS line was detected in previous observations \citep{hirota2001}. 
Recently, \citet{sakai2008-l1527} detected the lines of various 
carbon-chain molecules such as 
C$_{n}$H ($n$=4,5,6) and C$_{n}$H$_{2}$ ($n$=3,4,6) in this source. 
Because of their high excitation temperature ($>$10~K), 
\citet{sakai2008-l1527} proposed a different type of carbon-chain chemistry other than 
that in the cold dark cloud cores, and called it as Warm Carbon-Chain Chemistry (WCCC). 
The column density of HC$_{5}$N listed in Table \ref{tab-collte} is 3.5 times larger than 
that of \citet{sakai2008-l1527} due to a difference in the assumed 
excitation temperature (6.5~K); \citet{sakai2008-l1527} employed a higher excitation 
temperature of 12.3~K by assuming that the line mainly comes from a 
dense and warm part near the protostar. 
Since we observed a single transition for each molecule only toward 
the {\it{IRAS}} position, it is not possible from our data 
to discuss molecular distribution and their excitation condition in detail. 

\subsection{L1517D}
\label{sec-l1517d}

L1517D is a starless dense core in the Taurus molecular cloud \citep{benson1989}. 
We searched for the molecular lines around the reference position. 
The CCS line was detected only toward the (+80\arcsec, +80\arcsec) position, 
while the HC$_{3}$N and HC$_{5}$N lines were not detected toward all the 
observed positions, as shown in Figure \ref{fig-sp}(a). 
The NH$_{3}$ line was detected neither toward the 
reference position \citep{benson1989} nor the CCS position above, 
suggesting that L1517D would be one of the new candidates for 
CCPRs with an extremely low NH$_{3}$/CCS ratio. 
On the other hand, it would be possible that the weak CCS emission 
and the low NH$_{3}$/CCS ratio are due to the offset from the core center, 
as in the case of L1498 and L1544 where the CCS lines tend to be intense at 
the outer part of the dense cores \citep{kuiper1996, ohashi1999}. 
In L1517D, the position where the CCS line was detected corresponds to 
the edge of the {\it{Spitzer}} 160~$\mu$m 
continuum emission image with the 40\arcsec \ resolution \citep{kirk2007}, 
although the dense core is not detected in the higher 
resolution observations with SCUBA. 
Further high resolution mapping observations of both dust continuum emission 
and molecular lines are required to confirm whether L1517D is one of the CCPRs. 

\subsection{L1512}

L1512 is one of the well studied starless 
dense cores in the Taurus molecular cloud \citep{benson1989}. 
Since the CCS, HC$_{3}$N, and HC$_{5}$N lines were detected in the previous observations 
\citep{suzuki1992}, we carried out mapping observations of the CCS and HC$_{3}$N 
lines in the present study. The results are shown in Figure \ref{fig-maps}(c). 
The derived column densities of CCS and HC$_{3}$N are larger than 
those reported by \citet{suzuki1992} because the peak positions of CCS and HC$_{3}$N 
are significantly shifted from the reference position \citep{suzuki1992}. 
The position of the dust continuum emission is plotted in Figure \ref{fig-maps}(c) \citep{kirk2005}. 
We found that the integrated intensity maps of CCS and HC$_{3}$N do not show a central hole, 
which is consistent with the results by \citet{lee2003}. However, their peak 
positions are shifted 
from the dust continuum peak by (-24\arcsec, -17\arcsec) in right ascension and 
declination, respectively. In particular, the CCS emission 
traces the southwest edge of the dust continuum emission, possibly suggesting the 
depletion of CCS at the dust continuum peak. 

\subsection{L183(N)}

L183 is a well studied dark cloud, which is also known as L134N, 
and L183(N) is one of the ammonia cores in the filamentary 
dense molecular cloud L183 \citep{ungerechts1980, benson1989}. 
The estimated distance of L183 is about 140~pc, 
although some other results are also reported \citep{hilton1995}. 
The CCS lines were observed previously by \citet{benson1998}. 
We observed only toward one position, which 
corresponds to the position for the DNC and HN$^{13}$C observations 
\citep{hirota2001}. The column density of CCS derived for this position 
agrees well with the previous result \citep{benson1998}. 

\subsection{B68}

B68 is a dense starless core located in the Pipe Nebula 
at an estimated distance of 130-200~pc \citep{hilton1995, rathborne2008}. 
We detected the CCS, HC$_{3}$N, and HC$_{5}$N lines toward the reference 
position, although we could not carry out the mapping observations due to the limited 
observing time. 
The CCS and NH$_{3}$ maps were presented by \citet{lai2003}. 
In our study, the column density is derived toward the NH$_{3}$ peak, 
which is offset by 45\arcsec \ from the CCS peak \citep{lai2003}, 
so that the derived column density of CCS is different from 
that of \citet{lai2003}, by a factor of 2-3. 

\subsection{L492}
\label{sec-l492}

L492 is a Bok globule located in the Aquila rift at a distance of 200~pc \citep{lee1999}. 
This core was not listed in the NH$_{3}$ survey by \citet{benson1989}, while we included 
it in our survey because an intense CS line was detected toward this source \citep{lee2001}. 
In the initial survey observations conducted in 2002, 
we found that the CCS, HC$_{3}$N, and HC$_{5}$N lines toward the reference 
position of L492 are extremely strong. 
Therefore, we carried out follow-up mapping observations of L492 in 
various molecular lines. Further details were reported in \citet{hirota2006}. 

\subsection{L429-1}

L429-1 is also a Bok globule located in the Aquila rift at a distance of 
200~pc \citep{lee1999}. 
Similar to L492, this core was not included in the NH$_{3}$ survey by \citet{benson1989}, 
while we observed it based on the results of the CS survey by \citet{lee2001}. 
Because intense CCS and HC$_{3}$N lines were detected toward 
the reference position of L429-1, we carried out mapping observations 
of the CCS and HC$_{3}$N lines. 
The results are shown in Figure \ref{fig-maps}(d). 
The overall structures seen in the integrated intensity maps of 
CCS and HC$_{3}$N are similar, while the HC$_{3}$N map 
shows relatively complex clumpy structure. 
There seem to be three cores in our map. 
According to the dust continuum observations \citep{crapsi2005}, 
the CCS and HC$_{3}$N distributions are significantly 
shifted from the continuum peak position, 
although the mapped region of the dust continuum emission \citep{crapsi2005} 
is limited to the northwestern peak of the CCS and HC$_{3}$N cores. 
For much wider area, the mid-infrared absorption map is available. 
It shows an extended high extinction region 
toward east of the dust continuum peak \citep{bacmann2000}, which agrees well 
with the HC$_{3}$N and CCS maps. Another peak of HC$_{3}$N and the 
weak emission of the CCS lines can be seen toward 150\arcsec \ east 
of the dust continuum peak. We observed the NH$_{3}$ line only toward 
the reference position of L429-1, and hence, the NH$_{3}$/CCS ratio at 
the eastern peak of HC$_{3}$N, is not calculated. In spite of the 
intense HC$_{3}$N emission, the HC$_{5}$N line was not detected in L429-1. 

\subsection{L483}
\label{sec-l483}

L483 is a star-forming dense core in the Aquila rift at a distance of 200~pc 
\citep{hilton1995}. 
An intense CCS line was detected by \citet{benson1998}. 
We derived the column density which is consistent with that of \citet{benson1998}. 
We also detected intense HC$_{3}$N and HC$_{5}$N 
lines toward the reference position of L483 as shown in Figure \ref{fig-l483}. 
Then, we carried out mapping observations in the HC$_{3}$N and CCS lines. 
The results are shown in Figure \ref{fig-maps}(e). 
According to the previous observations of molecular lines 
and submillimeter continuum emission \citep{fuller1995}, 
L483 harbors a Class 0 protostar IRAS~18148-0440 
associated with molecular outflows. 
The CCS and HC$_{5}$N lines are quite strong toward 
the central protostar, although we cannot completely 
rule out the possibility that these centrally peaked structure 
would be an artifact due to insufficient spatial resolution, as seen in \citet{velusamy1995}. 
Extremely high abundances of carbon-chain molecules toward the Class 0 protostar 
are apparently similar to the L1527 core \citep{sakai2008-l1527}, and are 
indicative of WCCC \citep{sakai2009}. 
However, higher abundance of CCS than that in L1527 seems 
to be rather common characteristics for 
CCPRs such as L1495B, L1521B, L1521E and the cyanopolyyne 
peak of TMC-1 \citep{hirota2002, hirota2004}. 
Observations of higher excitation lines of carbon-chain molecules in L483 
would be crucial to distinguish whether L483 is categorized as WCCC or CCPR. 

\subsection{L530D}
\label{sec-l530d}

L530D is a dense core which is included in the survey of NH$_{3}$ by 
\citet{benson1989}. The estimated distance is 350~pc \citep{hilton1995}. 
We searched for the lines around the reference position of L530D, 
and found that the peak position of the CCS line is 
located at (+80\arcsec, +40\arcsec). 
However, the HC$_{3}$N and HC$_{5}$N lines were not detected toward this position, 
as shown in Figures \ref{fig-sp}(b). 
This position is offset by 85\arcsec \ northeast of the dust continuum peak 
\citep{kirk2005}. On the other hand, the NH$_{3}$ line was detected toward 
the dust continuum peak 
\citep{benson1989, kirk2005}, while it was not detected toward the CCS peak in our observations. 
Although the NH$_{3}$/CCS ratio at the CCS peak in L530D is as low as those of 
the other CCPRs, lower column densities of CCS and other 
carbon-chain molecules, the offsets of the CCS peak from the dust continuum emission, 
and the detection of NH$_{3}$ at the dust continuum peak would suggest that 
the CCS line traces the outer part of this core, similar to the case of L1517D. 
In fact, the column density of NH$_{3}$ toward the dust continuum peak 
is derived to be 7.9$\times$10$^{13}$~cm$^{-2}$ from the data of 
\citet{benson1989}, which results in the NH$_{3}$/CCS ratio toward the dust continuum 
peak 10 times larger than that toward the CCS peak. 
Molecular distributions in L530D seem to be similar to those in L1498 and L1544, 
where the NH$_{3}$ peak is coincident with the dust continuum and CCS is distributed 
in the surrounding region \citep{kuiper1996, ohashi1999}. 
However, we cannot totally rule out the possibility that L530D is a candidate for 
the CCPR because of a lack of complete mapping 
observations in the CCS and NH$_{3}$ lines. 

\subsection{L1147}
\label{sec-l1147}

L1147 is a dense core located in the Cepheus region at a distance of 
325~pc \citep{hilton1995}. This source is included in the survey of 
the NH$_{3}$ lines \citep{benson1989}. 
We searched for the CCS, HC$_{3}$N, and HC$_{5}$N lines around the reference 
position of L1147, and detected the CCS and HC$_{3}$N lines only toward the 
(0\arcsec, -80\arcsec) position, whose spectra are shown in Figures \ref{fig-sp}(c). 
The HC$_{5}$N line was not detected toward all the observed position. 
The NH$_{3}$ line was detected toward both the reference position 
\citep{benson1989} and the CCS peak. 
The position of the CCS peak is offset by only 20\arcsec \ toward south of the 
dust continuum peak of L1148, which is 1\arcmin \ south of L1147 
\citep{myerslinkebenson1983, kirk2005}, and hence, the dense core of 
L1147 found in the present survey can be regarded as L1148. 
The relatively low NH$_{3}$/CCS ratio toward the dust continuum peak of L1147 means 
that it could be a candidate for the CCPR. 

\subsection{L1172D}

L1172D was originally identified as a starless dense core 
in the survey of the NH$_{3}$ lines 
\citep{benson1989}, which is located in the Cepheus region 
at a distance of 288~pc \citep{hilton1995}. 
At first, the CCS and HC$_{3}$N lines were found to be intense toward the eastern 
part of the core of L1172D, and hence, the mapping region was extended toward 
1\arcmin-2\arcmin \ east from the reference position. 
Then the CCS and HC$_{3}$N emission peaks were found to be located toward 
L1172A, another star-forming dense core identified by \citet{benson1989}. 
There is a Class I protostar, {\it{IRAS}} 21017+6742, at the center of the core, 
which is known to be an engine of the molecular outflow \citep{myers1988}. 
The integrated intensity maps for HC$_{3}$N and CCS are shown in 
Figure \ref{fig-maps}(f). 
The HC$_{3}$N peak corresponds to the {\it{IRAS}} position, while CCS 
is significantly depleted toward the center of the HC$_{3}$N core. 
The shape and size of the HC$_{3}$N core is quite similar to that of 
NH$_{3}$ \citep{benson1989}. 
These results suggest that CCS is depleted as in 
the case of B335 \citep{velusamy1995}, while HC$_{3}$N is abundant 
toward the central part of the core. 

\subsection{L1172B}
\label{sec-l1172b}

L1172B is also identified as a starless dense core in the Cepheus region, 
which is observed in the survey of the NH$_{3}$ lines \citep{benson1989}. 
The observed spectra toward three positions in L1172B and the 
integrated intensity maps for HC$_{3}$N and CCS are shown in 
Figures \ref{fig-sp}(d)-(f) and \ref{fig-maps}(g), respectively. 
A dense core traced by the CCS and HC$_{3}$N lines shows an elongated 
structure extending from northeast to southwest, and 
consists of at least two clumps. 
The integrated intensity map of the NH$_{3}$ lines is also shown in 
Figure  \ref{fig-maps}(g) in thick contour lines. 
The NH$_{3}$ lines are detected only toward the western edge of 
the elongated molecular cloud core traced by the CCS and HC$_{3}$N lines. 
The NH$_{3}$/CCS abundance ratios are relatively low in the northeast clump, 
(+40\arcsec, +40\arcsec), 
and hence, this clump can be a candidate of CCPR. 
An {\it{IRAS}} point source, {\it{IRAS}} 21025+6801, is located at 
northwest of the core, although it is not clear whether this source is physically 
associated with the core of L1172B. 

\subsection{L183(S), L1696A, L1689N, L1157, L1155C, and L1228}

The results of the CCS observations have already been reported for these sources 
\citep{hirota2001}. 
We reanalyzed the HC$_{3}$N data for these sources, while we did not 
make further observations of the HC$_{5}$N lines. 

\section{Compilation of the samples of the CCS and NH$_{3}$ cores}
\label{sec-appendall}

As mentioned in Introduction, several survey observations of 
the CCS and NH$_{3}$ lines \citep{benson1989, fuente1990, suzuki1992, scappini1996, 
benson1998, turner1998, lai2000, hirota2001, degregorio-monsalvo2006, rathborne2008, 
rosolowsky2008, sakai2008, tatematsu2008} were carried out for the nearby molecular clouds. 
Here we compiled the results for CCS, as listed in Table \ref{tab-summary}. 
In this paper, we mainly focused on results of the observations of the CCS lines 
toward the nearby molecular cloud cores conducted with the single dish telescope 
in the 45~GHz band \citep{suzuki1992, benson1998, hirota2001} to compare with 
the present results, all of which are analyzed by almost the same methods. 

\subsection{CCS samples}

For the statistical study with uniform samples of nearby dark cloud cores, 
we excluded results of other surveys which were conducted in the different frequency 
bands and/or with interferometers 
\citep{scappini1996, lai2000, degregorio-monsalvo2006, rathborne2008, rosolowsky2008}, 
while we referred to the results by \citet{rathborne2008} and \citet{rosolowsky2008} 
for comparison of source-to-source variation of molecular abundances. 
We only employed dark cloud cores and low-mass star-forming regions 
in our sample as uniformly as possible. 
As a result, some of the sources in the high-mass star-forming molecular clouds 
in the Orion region \citep[e.g.,][]{tatematsu2008} were not included in 
our samples. 
We also excluded the different kind of samples such as translucent clouds 
\citep{turner1998} and distant infrared dark clouds \citep{sakai2008}. 
For a similar reason, we did not include the results of individual sources 
\citep[e.g.,][]{velusamy1995, kuiper1996, ohashi1999, lai2003, lee2003}, 
although these sources are included in the part of the survey observations. 
For the sources which are included in more than two literatures, 
we adopted one of the results as indicated in Table \ref{tab-summary} 
to avoid overlap of the samples. 
As a result, our sample consists of 90 nearby dark cloud cores 
as listed in Table \ref{tab-summary}. 

We compared the results of several sources which are observed both in 
our observations and previous ones, such as 
L1527, L1512, L1523, L183(N), 
L1719B, B68, L483, and L1172A (L1172D) 
\citep{suzuki1992, benson1998, hirota2001, lai2003}. 
As a result, we found that the difference in the derived column density is 
typically within 40\% but is a factor of 3 in the worst case, 
which is mostly due to the difference in the observed position 
\citep[e.g., in the case of L1512;][]{suzuki1992, benson1998}. 
These uncertainties are comparable to those attributed to the unknown 
excitation temperatures as already discussed in the main text. 

\subsection{NH$_{3}$ samples}

For the NH$_{3}$ lines, substantial amount of the sources listed in Table \ref{tab-summary} 
were observed by \citet{benson1989}. However, the column densities of NH$_{3}$ were reported 
for a part of the sources where the intense spectra were detected. Therefore, we calculated 
the column densities of NH$_{3}$ lines using the line parameters reported by \citet{benson1989}. 
The method is described by \citet{suzuki1992}, in which we assumed the excitation temperature of 6.5~K. 
If the NH$_{3}$ lines were not detected, we calculated the upper limits for the column densities 
by assuming the line width of 0.5 km s$^{-1}$. 

The data for the NH$_{3}$ lines are mostly taken from the systematic survey observations 
by \citet{benson1989} and \citet{suzuki1992}, while some individual observations are included 
\citep{ungerechts1980, mundy1990, bachiller1993, ladd1994, anglada1997, hirota2002, hirota2004}. 
Therefore, the uncertainties in the column densities of NH$_{3}$ might be affected by 
the difference in the observations and data analysis such as the telescope beam sizes and 
assumed and/or derived kinetic temperatures. 
However, we estimate that the uncertainties in the column densities of NH$_{3}$ is within 
a factor of 2 for most of the sources as discussed in \citet{suzuki1992}. 

\begin{deluxetable}{llllllrlrrrrl}
\rotate
\tabletypesize{\scriptsize}
\tablenum{A.1}
\tablewidth{0pt}
\tablecaption{Summary of Column Densities of CCS, HC$_{3}$N, HC$_{5}$N, and NH$_{3}$
 \label{tab-summary}}
\tablehead{
\colhead{} & \colhead{Source} & \colhead{} &
\colhead{} & \colhead{YSO({\it{IRAS}})} & \colhead{} & 
\colhead{Distance} & \colhead{} & \colhead{$N$[CCS]} & 
\colhead{$N$[HC$_{3}$N]} & \colhead{$N$[HC$_{5}$N]} & 
\colhead{$N$[NH$_{3}$]} & 
\colhead{} 
 \vspace{2mm} \\
\colhead{No} & \colhead{Name} & \colhead{RA(J2000)} &
\colhead{Dec(J2000)} & \colhead{Name} & \colhead{Region} & 
\colhead{(pc)} & \colhead{Ref.} & \colhead{(10$^{12}$~cm$^{-2}$)} & 
\colhead{(10$^{12}$~cm$^{-2}$)} & \colhead{(10$^{12}$~cm$^{-2}$)} & 
\colhead{(10$^{14}$~cm$^{-2}$)} & 
\colhead{Ref.} 
}
\startdata
       1  & L1455              & 03~27~40.3 & $+$30~13~03 &  L1455-FIR4  & Perseus        &  235    &   1      &   7.2      &  3.6       &   $<$3.8   &  7.3    &    1,1,1,1  \\
       2  & Per5               & 03~29~51.6 & $+$31~39~04 &  03267+3128  & Perseus        &  235    &   1      &   1.9      &   \nodata  &   \nodata  &  1.0    &    2,*,*,3  \\
       3  & B1                 & 03~33~16.3 & $+$31~07~51 &  03301+3057  & Perseus        &  235    &   1      &   13.8     &  8.5       &  8.8       &  11.2   &    1,1,1,1  \\
       4  & B5                 & 03~47~38.4 & $+$32~52~43 &  03445+3242  & Perseus        &  235    &   1      &   6.7      &  6.8       &   $<$2.8   &  2.3    &    1,1,1,1  \\
       5  & L1489             & 04~04~50.6 & $+$26~19~41 &  04016+2610  & Taurus         &  137    &   2      &   4.7      &  12.8      &  5.8       &  6.3    &    1,1,1,1  \\
       6  & L1498             & 04~10~51.5 & $+$25~09~58 &  \nodata     & Taurus         &  137    &   2      &   16.5     &  15.8      &  4.3       &  4.1    &    1,1,1,1  \\
       7  & L1495C/L1495N     & 04~13~45.7 & $+$28~13~14 &  04108+2803  & Taurus         &  137    &   2      &   17.2     &  13.8      &   $<$9.0   &  4.0    &    4,4,4,5  \\
       8  & L1495             & 04~14~12.1 & $+$28~09~30 &  04108+2803  & Taurus         &  137    &   2      &   $<$1.7   &  4.6       &   $<$3.2   &  5.1    &    1,1,1,1  \\
       9  & L1495D            & 04~14~18.2 & $+$28~15~52 &  \nodata     & Taurus         &  137    &   2      &   $<$2.2   &   $<$1.4   &   $<$4.9   & $<$1.0  &    4,4,4,5  \\
      10  & L1495B            & 04~15~36.5 & $+$28~47~06 &  \nodata     & Taurus         &  137    &   2      &   14.4     &  21        &  5.2       &  0.55   &    6,6,6,6  \\
      11  & L1506             & 04~18~31.1 & $+$25~19~25 &  \nodata     & Taurus         &  137    &   2      &   $<$2.8   &   $<$1.7   &   $<$6.2   & $<$0.5  &    4,4,4,5  \\
      12  & L1521C            & 04~19~19.2 & $+$27~16~29 &  \nodata     & Taurus         &  137    &   2      &   $<$1.1   &   $<$0.7   &   $<$2.5   & $<$0.3  &    1,1,1,1  \\
      13  & L1521B            & 04~24~12.7 & $+$26~36~53 &  \nodata     & Taurus         &  137    &   2      &   36       &  41        &  12        &  1.26   &    6,6,6,6  \\
      14  & L1400B            & 04~24~37.3 & $+$55~02~33 &  \nodata     & other          &  170    &   3      &   5.1      &  5.0       &   $<$5.3   &  0.62   &    4,4,4,4  \\
      15  & L1521A            & 04~26~43.9 & $+$26~15~48 &  \nodata     & Taurus         &  137    &   2      &   $<$0.9   &   $<$0.7   &   $<$2.0   & $<$0.3  &    1,1,1,1  \\
      16  & L1521D            & 04~27~46.5 & $+$26~17~52 &  \nodata     & Taurus         &  137    &   2      &   3.6      &  1.2       &   $<$2.0   &  1.9    &    1,1,1,1  \\
      17  & L1400E            & 04~28~28.5 & $+$54~47~36 &  \nodata     & other          &  170    &   3      &   $<$2.6   &   $<$1.6   &   $<$5.6   & $<$0.82 &    4,4,4,5  \\
      18  & L1521E            & 04~29~16.5 & $+$26~13~50 &  \nodata     & Taurus         &  137    &   2      &   28       &  23        &  4.6       &  0.73   &    7,7,7,7  \\
      19  & L1400F            & 04~29~51.4 & $+$54~14~19 &  \nodata     & other          &  170    &   3      &   $<$2.4   &   $<$1.6   &   $<$5.1   &  0.84   &    4,4,4,5  \\
      20  & L1400K            & 04~30~52.1 & $+$54~51~55 &  \nodata     & other          &  170    &   3      &   3.6      &  6.4       &   $<$2.3   &  1.8    &    1,1,1,1  \\
      21  & L1551A            & 04~30~58.1 & $+$18~17~10 &  \nodata     & Taurus         &  137    &   2      &   $<$2.6   &   $<$1.5   &   $<$5.5   & $<$0.98 &    4,4,4,5  \\
      22  & L1551             & 04~31~30.0 & $+$18~12~30 &  04287+1806  & Taurus         &  137    &   2      &   $<$2.3   &   $<$1.4   &   $<$5.0   & $<$0.72 &    4,4,4,5  \\
      23  & L1551S            & 04~31~33.9 & $+$18~08~05 &  04287+1801  & Taurus         &  137    &   2      &   $<$2.2   &  6.2       &   $<$1.8   &  5.4    &    1,1,1,1  \\
      24  & TMC-2A            & 04~31~55.9 & $+$24~32~49 &  04292+2422  & Taurus         &  137    &   2      &   3.2      &  3.7       &   $<$1.6   &  10.7   &    1,1,1,1  \\
      25  & L1445             & 04~32~07.0 & $+$46~37~23 &  \nodata     & other          & \nodata &  \nodata &   $<$2.3   &   $<$1.6   &   $<$5.3   & \nodata &    4,4,4,*  \\
      26  & TMC-2             & 04~32~44.8 & $+$24~25~12 &  \nodata     & Taurus         &  137    &   2      &   25       &  33        &  6.7       &  6.1    &    1,1,1,1  \\
      27  & L1536B            & 04~33~25.6 & $+$22~43~26 &  \nodata     & Taurus         &  137    &   2      &   6.6      &  5.7       &   $<$2.6   &  4.9    &    1,1,1,1  \\
      28  & L1527             & 04~39~53.6 & $+$26~03~05 &  04368+2557  & Taurus         &  137    &   2      &   5.1      &  47        &  18.9      &  5.0    &    8,4,4,5  \\
      29  & TMC-1(NH$_{3}$)   & 04~41~23.0 & $+$25~48~13 &  04381+2540  & Taurus         &  137    &   2      &   18.8     &  61        &  10.0      &  5.2    &    1,1,1,1  \\
      30  & TMC-1C            & 04~41~34.3 & $+$26~00~43 &  04385+2550  & Taurus         &  137    &   2      &   31       &  28        &  9.4       &  5.7    &    1,1,1,1  \\
      31  & TMC-1(CP)         & 04~41~42.5 & $+$25~40~42 &  \nodata     & Taurus         &  137    &   2      &   66       &  171       &  56        &  1.9    &    1,1,1,1  \\
      32  & L1517C            & 04~54~42.8 & $+$30~34~48 &  \nodata     & Taurus         &  137    &   2      &   $<$1.1   &   $<$0.7   &   $<$1.9   & $<$0.3  &    1,1,1,1  \\
      33  & L1517A            & 04~55~06.3 & $+$30~33~40 &  \nodata     & Taurus         &  137    &   2      &   $<$0.9   &   $<$0.7   &   $<$1.9   &  0.5    &    1,1,1,1  \\
      34  & L1517B            & 04~55~18.8 & $+$30~38~04 &  \nodata     & Taurus         &  137    &   2      &   8.6      &  9.3       &  4.3       &  7.0    &    1,1,1,1  \\
      35  & L1517D            & 04~55~54.4 & $+$30~40~05 &  \nodata     & Taurus         &  137    &   2      &   3.7      &   $<$1.1   &   $<$3.7   & $<$0.13 &    4,4,4,4  \\
      36  & L1512             & 05~04~06.5 & $+$32~43~09 &  \nodata     & Taurus         &  137    &   2      &   9.0      &  25        &   $<$2.1   &  7.0    &    4,4,1,1  \\
      37  & L1544             & 05~04~15.2 & $+$25~11~08 &  \nodata     & Taurus         &  137    &   2      &   18.0     &  17.8      &  7.8       &  2.7    &    1,1,1,1  \\
      38  & L1523             & 05~06~22.9 & $+$31~41~19 &  \nodata     & Taurus         &  137    &   2      &   $<$2.2   &   $<$1.4   &   $<$4.8   & $<$1.06 &    4,4,4,5  \\
      39  & L1778A            & 15~39~27.5 & $-$07~10~08 &  \nodata     & other          &  100    &   3      &   $<$3.0   &   $<$1.6   &   $<$6.3   & $<$1.03 &    4,4,4,5  \\
      40  & L183(N)           & 15~54~09.2 & $-$02~49~42 &  \nodata     & other          &  140    &   3      &   3.6      &  13.7      &   $<$7.3   &  16     &    4,4,4,9  \\
      41  & L183(S)           & 15~54~09.2 & $-$02~51~39 &  \nodata     & other          &  140    &   3      &   7.2      &  4.1       &   \nodata  &  32     &    8,4,*,9  \\
      42  & L1721             & 16~14~28.2 & $-$18~54~44 &  \nodata     & Ophiuchus      &  120    &   4      &   $<$3.1   &   $<$1.8   &   $<$7.0   & $<$0.81 &    4,4,4,5  \\
      43  & L1719B            & 16~22~12.4 & $-$19~38~41 &  \nodata     & Ophiuchus      &  120    &   4      &   $<$3.3   &   $<$1.8   &   $<$7.0   & $<$0.87 &    4,4,4,5  \\
      44  & L1687             & 16~23~01.6 & $-$22~53~41 &  \nodata     & Ophiuchus      &  120    &   4      &   $<$1.5   &   $<$0.8   &   $<$2.6   & $<$0.3  &    1,1,1,1  \\
      45  & L1681A            & 16~26~26.6 & $-$24~33~16 &  \nodata     & Ophiuchus      &  120    &   4      &   $<$2.4   &   $<$1.0   &   $<$3.4   & $<$0.4  &    1,1,1,1  \\
      46  & L1681B            & 16~27~26.9 & $-$24~44~27 &  16244-2432  & Ophiuchus      &  120    &   4      &   $<$1.7   &   $<$1.0   &   $<$3.2   &  1.2    &    1,1,1,1  \\
      47  & L1690             & 16~27~46.4 & $-$24~16~59 &  \nodata     & Ophiuchus      &  120    &   4      &   $<$3.1   &   $<$1.8   &   $<$7.0   & $<$0.72 &    4,4,4,5  \\
      48  & L1696A            & 16~28~31.5 & $-$24~19~08 &  \nodata     & Ophiuchus      &  120    &   4      &   $<$2.7   &   $<$2.1   &   $<$2.0   &  5.0    &    8,4,10,5  \\
      49  & L1696B            & 16~29~03.0 & $-$24~21~33 &  \nodata     & Ophiuchus      &  120    &   4      &   $<$1.3   &   $<$0.8   &   $<$2.3   & $<$0.3  &    1,1,1,1  \\
      50  & L1709A            & 16~30~50.8 & $-$23~41~03 &  \nodata     & Ophiuchus      &  120    &   4      &   $<$3.3   &   $<$1.8   &   $<$7.2   &  0.99   &    4,4,4,5  \\
      51  & L1709B            & 16~31~39.6 & $-$24~01~23 &  16285-2356  & Ophiuchus      &  120    &   4      &   3.5      &  5.9       &   $<$2.3   &  1.6    &    1,1,1,1  \\
      52  & L1689A            & 16~32~12.6 & $-$25~02~53 &  \nodata     & Ophiuchus      &  120    &   4      &   $<$3.2   &   $<$1.7   &   $<$4.8   &  1.7    &    1,1,1,1  \\
      53  & L1689N            & 16~32~22.8 & $-$24~28~33 &  16293-2422  & Ophiuchus      &  120    &   4      &   5.8      &  7.5       &   \nodata  &  20     &    8,4,*,11  \\
      54  & $\rho$-Oph-E      & 16~32~29.5 & $-$24~28~13 &  16293-2422  & Ophiuchus      &  120    &   4      &   $<$2.2   &  4.3       &   $<$2.8   &  10.6   &    1,1,1,1  \\
      55  & L1709C            & 16~33~53.4 & $-$23~38~32 &  \nodata     & Ophiuchus      &  120    &   4      &   $<$3.3   &   $<$1.7   &   $<$6.9   &  1.14   &    4,4,4,5  \\
      56  & L1709             & 16~34~36.0 & $-$23~43~11 &  \nodata     & Ophiuchus      &  120    &   4      &   $<$1.5   &   $<$0.9   &   $<$2.7   &  0.7    &    1,1,1,1  \\
      57  & L43E              & 16~34~39.8 & $-$15~47~00 &  16316-1540  & Ophiuchus      &  120    &   4      &   4.5      &  3.7       &   $<$1.8   &  10.1   &    1,1,1,1  \\
      58  & L1689B            & 16~34~42.1 & $-$24~36~11 &  \nodata     & Ophiuchus      &  120    &   4      &   $<$1.1   &   $<$0.8   &   $<$1.9   &  0.8    &    1,1,1,1  \\
      59  & L260              & 16~47~06.7 & $-$09~35~21 &  16442-0930  & Ophiuchus      &  120    &   4      &   1.6      &   \nodata  &   $<$7.9   &  15.8   &    2,*,10,5    \\
      60  & L158              & 16~47~23.2 & $-$13~59~21 &  16445-1352  & Ophiuchus      &  120    &   4      &   $<$2.8   &   $<$1.7   &   $<$7.0   &  4.0    &    4,4,4,5  \\
      61  & L191              & 16~47~29.3 & $-$12~28~38 &  \nodata     & Ophiuchus      &  120    &   4      &   $<$2.8   &   $<$1.7   &   $<$6.4   &  0.71   &    4,4,4,5  \\
      62  & L204F             & 16~47~48.4 & $-$11~56~56 &  \nodata     & Ophiuchus      &  120    &   4      &   $<$2.9   &   $<$1.7   &   $<$6.1   &  1.43   &    4,4,4,5  \\
      63  & L234A             & 16~48~06.9 & $-$10~51~48 &  16451-1045  & Ophiuchus      &  120    &   4      &   $<$3.3   &   \nodata  &   $<$3.2   &  5.0    &    2,*,10,5    \\
      64  & L234E             & 16~48~08.6 & $-$10~56~58 &  16451-1054  & Ophiuchus      &  120    &   4      &   2.1      &   \nodata  &   \nodata  &  2.5    &    2,*,*,5    \\
      65  & L63               & 16~50~15.5 & $-$18~06~06 &  \nodata     & Ophiuchus      &  120    &   4      &   2.3      &  3.9       &   $<$5.1   &  7.9    &    2,1,1,1  \\
      66  & B68               & 17~22~38.8 & $-$23~50~02 &  \nodata     & other          &  200    &   3      &   3.0      &  18.9      &  7.8       &  2.5    &    4,4,4,5  \\
      67  & L492              & 18~15~46.1 & $-$03~46~13 &  \nodata     & Aquila         &  200    &   6      &   53       &  172       &  41        &  3.4    &   12,12,12,12  \\
      68  & L429-1            & 18~17~05.6 & $-$08~13~30 &  \nodata     & Aquila         &  200    &   6      &   14.2     &  7.5       &   $<$6.4   &  10.4   &    4,4,4,4  \\
      69  & L483              & 18~17~29.7 & $-$04~39~38 &  18148-0440  & Aquila         &  200    &   3      &   12.3     &  31        &  17.6      &  14.9   &    4,4,4,4  \\
      70  & L530H             & 18~49~28.5 & $-$04~57~40 &  \nodata     & Aquila         &  350    &   3      &   $<$3.8   &   $<$2.0   &   $<$8.4   & $<$0.65 &    4,4,4,5  \\
      71  & L530D             & 18~50~02.7 & $-$04~49~09 &  \nodata     & Aquila         &  350    &   3      &   4.1      &   $<$1.5   &   $<$5.4   & $<$0.09 &    4,4,4,4  \\
      72  & B133              & 19~06~06.9 & $-$06~52~41 &  \nodata     & Aquila         &  400    &   3      &   7        &   \nodata  &   \nodata  &  2.89   &    2,*,*,5  \\
      73  & L723              & 19~17~53.9 & $+$19~12~19 &  19156+1906  & other          &  300    &   3      &   $<$2.8   &   $<$1.3   &   $<$2.8   &  1.6    &    1,1,1,1  \\
      74  & L778              & 19~26~32.6 & $+$23~58~42 &  19244+2352  & other          &  200    &   3      &   4.5      &  4.3       &  6.1       &  2.8    &    1,1,1,1  \\
      75  & B335              & 19~36~59.0 & $+$07~33~47 &  19345+0727  & other          &  250    &   3      &   2.9      &  4.2       &   $<$2.2   &  1.3    &    2,1,1,1  \\
      76  & L1152             & 20~35~50.3 & $+$67~54~22 &  20353+6742  & Cepheus        &  325    &   3      &   $<$4.2   &   \nodata  &   \nodata  &  7.9    &    2,*,*,5  \\
      77  & L1157             & 20~39~06.6 & $+$68~02~13 &  20386+6751  & Cepheus        &  325    &   3      &   $<$2.3   &  1.64      &   \nodata  &  5.2    &    8,4,*,13  \\
      78  & L1147             & 20~40~31.9 & $+$67~20~25 &  \nodata     & Cepheus        &  325    &   3      &   5.5      &  4.5       &   $<$5.9   &  0.08   &    4,4,4,4  \\
      79  & L1155H            & 20~43~06.5 & $+$67~46~26 &  \nodata     & Cepheus        &  325    &   3      &   $<$2.4   &   $<$1.6   &   $<$5.4   & $<$0.82 &    4,4,4,5  \\
      80  & L1155C            & 20~43~30.0 & $+$67~52~42 &  \nodata     & Cepheus        &  325    &   3      &   5.2      &  2.4       &   \nodata  &  5.0    &     8,4,*,5    \\
      81  & L1155D            & 20~43~49.8 & $+$67~36~29 &  \nodata     & Cepheus        &  325    &   3      &   $<$2.8   &   $<$1.6   &   $<$5.7   & $<$0.34 &    4,4,4,5  \\
      82  & L1082A            & 20~53~30.2 & $+$60~14~46 &  20520+6003  & Cepheus        &  440    &   3      &   2.6      &   3.3      &   $<$1.7   & 1.5     &    2,2,2,2  \\
      83  & L1082B            & 20~53~52.2 & $+$60~11~18 &  20526+5958  & Cepheus        &  440    &   3      &   $<$1.3   &   $<$0.7   &   $<$2.2   & 0.6     &    2,2,2,2 \\
      84  & L1228             & 20~59~30.6 & $+$78~22~49 &  21004+7811  & Cepheus        &  150    &   3      &   $<$2.4   &  2.3       &   \nodata  &  1.74   &     8,4,*,5    \\
      85  & L1172A/L1172D     & 21~02~23.2 & $+$67~54~35 &  21017+6742  & Cepheus        &  288    &   3      &   4.9      &  21        &  14.4      &  7.9    &    4,4,4,5  \\
      86  & L1172B(-120,-40)  & 21~03~10.6 & $+$68~11~17 &  \nodata     & Cepheus        &  288    &   3      &   4.0      &  4.4       &   $<$5.2   &   0.81  &    4,4,4,4  \\
      87  & L1172B(-40,-40)   & 21~03~24.9 & $+$68~11~18 &  \nodata     & Cepheus        &  288    &   3      &   4.6      &  6.4       &   $<$5.0   & $<$0.13 &    4,4,4,4  \\
      88  & L1172B(+40,+40)   & 21~03~39.1 & $+$68~12~39 &  \nodata     & Cepheus        &  288    &   3      &   6.4      &  4.2       &   $<$4.8   & $<$0.13 &    4,4,4,4  \\
      89  & B361              & 21~12~13.2 & $+$47~24~24 &  21106+4712  & other          &  350    &   3      &   $<$1.9   &   $<$0.7   &   $<$3.6   &  0.7    &    1,1,1,1  \\
      90  & L1251E            & 22~39~17.7 & $+$75~11~29 &  22376+7455  & Cepheus        &  300    &   3      &   6        &   \nodata  &   \nodata  &  3.1    &    2,*,*,14  \\
\enddata
\tablecomments{References for distances --- 
1: \citet{hirota2008}; 2: \citet{torres2007}; 3: \citet{hilton1995}; 
4: \citet{loinard2008}; 5: \citet{lee1999}}
\tablecomments{References for column densities --- 
1: \citet{suzuki1992}; 2: \citet{benson1998}; 3: \citet{ladd1994}; 
4: Present study; 5: \citet{benson1989}; 6: \citet{hirota2004}; 
7: \citet{hirota2002}; 8: \citet{hirota2001}; 9: \citet{ungerechts1980}; 
10: \citet{benson1983}; 11: \citet{mundy1990}; 12: \citet{hirota2006}; 
13: \citet{bachiller1993}; 14: \citet{anglada1997}}
\end{deluxetable}


\newpage

\begin{thebibliography}{}
\bibitem[Aikawa et al.(2005)]{aikawa2005}  Aikawa, Y., Herbst, E., 
  Roberts, H., \& Caselli, P. 2005, ApJ, 620, 330
\bibitem[Aikawa et al.(2003)]{aikawa2003}  Aikawa, Y., Ohashi, N., \& 
  Herbst, E. 2003, ApJ, 593, 906
\bibitem[Aikawa et al.(2001)]{aikawa2001} Aikawa, Y., Ohashi, N., 
  Inutsuka, S., Herbst, E., \& Takakuwa, S. 2001, ApJ, 552, 639
\bibitem[Alexander et al.(1976)]{alexander1976} Alexander, A. J., 
  Kroto, H. W., \& Walton, D. R. M. 1976, J. Mol. Spectrosc., 62, 175 
\bibitem[Anglada et al.(1997)]{anglada1997} Anglada, G., Sepulveda, I., \& Gomez, J. F. 
  1997, A\&AS, 121, 255
\bibitem[Bachiller et al.(1993)]{bachiller1993} Bachiller, R., 
  Martin-P\'\i ntado, J., \& Fuente, A. 1993, ApJ, 417, L45
\bibitem[Bacmann et al.(2000)]{bacmann2000}
  Bacmann, A., Andr\'e, P., Puget, J.-L., Abergel, A., Bontemps, S., 
  \& Ward-Thompson, D. 2000, A\&A, 361, 555
\bibitem[Benson et al.(1998)]{benson1998} 
  Benson, P. J., Caselli, P., \& Myers, P. C. 1998, ApJ, 506, 743
\bibitem[Benson \& Myers(1983)]{benson1983} 
  Benson, P. J., \& Myers, P. C. 1983, ApJ, 270, 589
\bibitem[Benson \& Myers(1989)]{benson1989} 
  Benson, P. J., \& Myers, P. C. 1989, ApJS, 71, 89
\bibitem[Bergin \& Langer(1997)]{bergin1997} 
  Bergin, E. A. \& Langer, W. D. 1997, ApJ, 486, 316
\bibitem[Brooke et al.(2007)]{brooke2007} Brooke, T. Y. et al. 2007, ApJ, 655, 364
\bibitem[Cohen \& Poynter(1974)]{cohen1974} Cohen, E. A., \& Poynter, R. L. 1974, 
  J. Mol. Spectrosc., 53, 131
\bibitem[Crapsi et al.(2005)]{crapsi2005}
  Crapsi, A., Caselli, P., Walmsley, C. M., Myers, P. C., Tafalla, M., 
  Lee, C. W., \& Bourke, T. L. 2005, ApJ, 619, 379
\bibitem[de Gregorio-Monsalvo et al.(2006)]{degregorio-monsalvo2006}
  de Gregorio-Monsalvo, I., G\'omez, J. F., Su\'arez, O., 
  Kuiper, T. B. H., Rodr\'\i guez, L. F., \& Jim\'enez-Bail\'on, E. 2006, ApJ, 642, 319
\bibitem[Enoch et al.(2006)]{enoch2006} Enoch, M. L. et al. 2006, ApJ, 638, 293
\bibitem[Fuente et al.(1990)]{fuente1990} Fuente, A., Cernicharo, J., 
  Barcia, A., \& Gomez-Gonz\'alez, J. 1990, A\&A, 231, 151
\bibitem[Fuller et al.(1995)]{fuller1995} 
  Fuller, G. A. Lada, E. A., Masson, C. R., \& Myers, P. C. 1995, ApJ, 453, 754
\bibitem[Hilton \& Lahulla(1995)]{hilton1995}
  Hilton, J. \& Lahulla, J. F. 1995, A\&AS, 113, 325
\bibitem[Hirahara et al.(1992)]{hirahara1992} 
  Hirahara, Y. et al. 1992, ApJ, 394, 539
\bibitem[Hirota et al.(2001)]{hirota2001}
  Hirota, T., Ikeda, M., \& Yamamoto, S. 2001, ApJ, 547, 814
\bibitem[Hirota et al.(2002)]{hirota2002}
  Hirota, T., Ito, T., \& Yamamoto, S. 2002, ApJ, 565, 359
\bibitem[Hirota et al.(2004)]{hirota2004}
  Hirota, T., Maezawa, H., \& Yamamoto, S. 2004, ApJ, 617, 399
\bibitem[Hirota \& Yamamoto(2006)]{hirota2006}
  Hirota, T., \& Yamamoto, S. 2006, ApJ, 646, 258
\bibitem[Hirota et al.(2008)]{hirota2008} Hirota, T. et al. 2008, PASJ, 60, 37
\bibitem[Jijina et al.(1999)]{jijina1999} Jijina, J., Myers, P. C., \& Adams, F. C. 
  1999, ApJS, 125, 161
\bibitem[J\o rgensen et al.(2008)]{jorgensen2008} J\o rgensen, J. K., Johnstone, D., 
   Kirk, H., Myers, P. C., Allen, L. E., \& Shirley, Y. L. 2008, ApJ, 683, 822
\bibitem[Kirk et al.(2005)]{kirk2005} Kirk, J. M., 
  Ward-Thompson, D., \& Andr\'e, P. 2005, MNRAS, 360, 1506
\bibitem[Kirk et al.(2007)]{kirk2007} Kirk, J. M., 
  Ward-Thompson, D., \& Andr\'e, P. 2007, MNRAS, 375, 843
\bibitem[Kuiper et al.(1996)]{kuiper1996}
  Kuiper, T. B. H., Langer, W. D., \& Velusamy, T. 1996, ApJ, 468, 761
\bibitem[Ladd et al.(1994)]{ladd1994}
  Ladd, E. F., Myers, P. C., \& Goodman, A. A. 1994, ApJ, 433, 117
\bibitem[Lafferty \& Lovas(1978)]{lafferty1978} 
  Lafferty, W. J. \& Lovas, F. J. 1978, J. Phys. Chem. Ref. Data, 7, 441
\bibitem[Lai \& Crutcher(2000)]{lai2000} 
  Lai, S.-P. \& Crutcher, R. M. 2000, ApJS, 128, 271
\bibitem[Lai et al.(2003)]{lai2003} 
  Lai, S.-P., Velusamy, T., Langer, W. D., \& Kuiper, T. B. H. 2003, AJ, 126, 311
\bibitem[Lee \& Myers(1999)]{lee1999} Lee, C. W., \& Myers, P. C. 1999, ApJS, 123, 233
\bibitem[Lee et al. (2001)]{lee2001} 
  Lee, C. W., Myers, P. C., \& Tafalla, M. 2001, ApJS, 136, 703
\bibitem[Lee et al.(2003)]{lee2003}
  Lee, J. -E., Evans, N. J. II., Shirley, Y. L., \& Tatematsu, K. 
  2003, ApJ, 583, 789
\bibitem[Loinard et al.(2008)]{loinard2008} 
  Loinard, L., Torres, R. M., Mioduszewski, A. J., \& Rodr\'\i guez, L. F.
  2008, ApJ, 675, L29
\bibitem[Mundy et al.(1990)]{mundy1990}
  Mundy, L. G., Wootten, H. A. Wilking, B. A. 1990, ApJ, 352, 159
\bibitem[Murakami(1990)]{murakami1990} Murakami, A. 1990, ApJ, 357, 288
\bibitem[Myers et al.(1988)]{myers1988}
  Myers, P. C., Heyer, M., Snell, R. L., \& Goldsmith, P. F. 1988, ApJ, 324, 907
\bibitem[Myers et al.(1983)]{myerslinkebenson1983} 
  Myers, P. C., Linke, R. A., \& Benson, P. J. 1983, ApJ, 264, 517
\bibitem[Ohashi et al.(1999)]{ohashi1999} 
  Ohashi, N., Lee, S. W., Wilner, D. J., \& Hayashi, M. 1999, ApJL, 518, L41
\bibitem[Ohishi \& Kaifu(1998)]{ohishi1998} 
  Ohishi, M., \& Kaifu, N. 1998, in Chemistry and Physics of Molecules and
  Grains in Space ( London: Royal Society of Chemistry), 205
\bibitem[Rathborne et al.(2008)]{rathborne2008} Rathborne, J. M., Lada, C. J., 
  Muenchi, A. A., Alves, J. F., \& Lombardi, M. 2008, ApJS, 174, 396
\bibitem[Rosolowsky et al.(2008)]{rosolowsky2008} Rosolowsky, E. W., 
  Pineda, J. E., Foster, J.  B., Borkin, M. A., Kauffmann, J., Caselli, P., 
  Myers, P. C., \& Goodman, A. A. 2008, ApJS, 175, 509
\bibitem[Sakai et al.(2009)]{sakai2009} Sakai, N., Sakai, T., Hirota, T., 
  Burton, M. G., \& Yamamoto, S. 2009, ApJ, 697, 769
\bibitem[Sakai et al.(2008a)]{sakai2008-l1527} Sakai, N., Sakai, T., Hirota, T., 
  \& Yamamoto, S. 2008a, ApJ, 672, 371
\bibitem[Sakai et al.(2008b)]{sakai2008} Sakai, T., Sakai, N., 
  Kamegai, K., Hirota, T., Yamaguchi, N., Shiba, S., \& Yamamoto, S. 2008b, ApJ, 678, 1049
\bibitem[Scappini \& Codella(1996)]{scappini1996}
  Scappini, F. \& Codella, C. 1996, MNRAS, 282, 587
\bibitem[Shirley et al.(2005)]{shirley2005}
  Shirley, Y. L., Nordhaus, M. K., Grcevich, J. M., Evans, N. J. II., 
  Rawlings, J. M. C., \& Tatematsu, K. 2005, ApJ, 632, 982
\bibitem[Suzuki et al.(1992)]{suzuki1992} 
  Suzuki, H., Yamamoto, S., Ohishi, M., Kaifu, N., Ishikawa, S., 
  Hirahara, Y., \& Takano, S. 1992, ApJ, 392, 551
\bibitem[Tachihara et al.(2000)]{tachihara2000}
  Tachihara, K., Mizuno, A., \& Fukui, Y. 2000, ApJ, 528, 817 
\bibitem[Tafalla \& Santiago(2004)]{tafalla2004}
  Tafalla, M., \& Santiago, J. 2004, A\&A, 414, L53
\bibitem[Tafalla et al.(2006)]{tafalla2006}
  Tafalla, M.\, Santiago-Garc\i\'ia, J., Myers, P. C., Caselli, P., Walmsley, C. M., 
  \&  Crapsi, A. 2006, A\&A, 455, 577
\bibitem[Tatematsu et al.(2008)]{tatematsu2008} Tatematsu, K., Kandori, R., 
   Umemoto, T., \& Sekimoto, Y. 2008, PASJ, 60, 407
\bibitem[Torres et al.(2007)]{torres2007} 
  Torres, R. M., Loinard, L., Mioduszewski, A. J., \& Rodr\'\i guez, L. F.
  2007, ApJ, 671, 1813
\bibitem[Turner et al.(1998)]{turner1998}
  Turner, B. E., Lee, H. -H., \& Herbst, E. 1998, ApJS, 115, 91
\bibitem[Ungerechts et al.(1980)]{ungerechts1980}
  Ungerechts, H., Walmsley, C. M., \& Winnewisser, G. 1980, A\&A, 88, 259
\bibitem[Velusamy et al.(1995)]{velusamy1995} Velusamy, T., Kuiper, T. B. H., \& 
  Langer, W. D. 1995, ApJ, 451, L75  
\bibitem[Ward-Thompson et al.(2007)]{ward-thompson2007}
  Ward-Thompson, D., Andr\'e, P., Crutcher, R., Johnstone, D., Onishi, T., \& Wilson, C. 
  2007, in Protostars and Planets V, eds. B. Reipurth, D. Jewitt, \& K. Keil (Tucson: 
  University of Arizona Press), 33
\bibitem[Yamamoto et al.(1990)]{yamamoto1990} Yamamoto, S., Shuji, S., Kawaguchi, S., 
  Chikada, Y., Suzuki H., Kaifu, N., Ishikawa, S., 
  \& Ohishi, M. 1990, ApJ, 361, 318 
\end{thebibliography}
\end{document}